\documentclass[amssymb,twocolumn,floats,amsmath,superscriptaddress]{revtex4-1}
\usepackage{graphicx}
\usepackage{textcomp}
\usepackage{float}
\usepackage{xcolor}
\usepackage[normalem]{ulem}
\usepackage{upgreek}
\usepackage[english]{babel}
\usepackage[utf8]{inputenc}
\usepackage{tikz}
\usepackage{braket}
\usepackage{cleveref}
\usepackage{amssymb}
\usepackage[colorinlistoftodos, color=green!40, prependcaption]{todonotes}

\usepackage{amsthm}
\usepackage{mathtools}
\usepackage{physics}
\usepackage{xcolor}
\usepackage{graphicx}
\usepackage[left=23mm,right=13mm,top=35mm,columnsep=15pt]{geometry} 
\usepackage{adjustbox}
\usepackage{placeins}
\usepackage[T1]{fontenc}
\usepackage{csquotes}



\def \IFT{Institute of Theoretical Physics, Faculty of Physics, University of Warsaw, Pasteura St. 5, 02-093 Warsaw, Poland}
\def \PW{Department of Semiconductor Materials Engineering Faculty of Fundamental Problems of Technology Wrocław University of Science and Technology Wybrzeże Wyspiańskiego 27, 50-370 Wrocław, Poland}
\def \AC{II. Institute of Physics B and JARA-FIT, RWTH-Aachen University, 52074 Aachen, Germany}
\def \Haifa{Schulich Faculty of Chemistry, Solid State Institute, Russell Berrie Nanotechnology Institute, Technion, Haifa 3200003, Israel}
\def \HaifaMat{Department of Materials Science and Engineering, Technion – Israel Institute of Technology, Haifa 3200003, Israel}
\def \FZJ{Forschungszentrum Jülich, Peter Grünberg Institute (PGI-6),
52425 Jülich, Germany}

\def \Trieste{Physics Department, University of Trieste, Trieste, 34127 Italy}
\def \Elettra{Elettra – Sincrotrone Trieste S.C.p.A., S.S. 14 km 163.5, Trieste, 34149 Italy}
\def \PNNL{Physical and Computational Sciences, Directorate and Institute for Integrated Catalysis, Pacific Northwest National Laboratory, Richland, WA 99354, USA}
\def \ac{2nd Institute of Physics and JARA-FIT, RWTH-Aachen University, Aachen 52074, Germany}

\begin{document}

\title{Probing the band structure of the strongly correlated antiferromagnet NiPS$_3$ across its phase transition}

\author{Benjamin Pestka$^\dag$}\affiliation{\AC}
\author{Biplab Bhattacharyya$^\dag$}\affiliation{\AC}
\author{Miłosz Rybak$^\dag$}\affiliation{\PW}
\author{Jeff Strasdas}\affiliation{\AC}
\author{Adam K. Budniak}\affiliation{\Haifa}
\author{Adi Harchol}\affiliation{\Haifa}
\author{Marcus Liebmann}\affiliation{\AC}
\author{Niklas Leuth}\affiliation{\AC}
\author{Honey Boban}\affiliation{\FZJ}
\author{Vitaliy Feyer}\affiliation{\FZJ}
\author{Iulia Cojocariu}\affiliation{\FZJ}\affiliation{\Trieste}\affiliation{\Elettra}
\author{Daniel Baranowski}\affiliation{\FZJ}\affiliation{\PNNL}
\author{Simone Mearini}\affiliation{\FZJ}
\author{Lutz Waldecker}\affiliation{\ac}
\author{Bernd Beschoten}\affiliation{\ac}
\author{Christoph Stampfer}\affiliation{\ac}
\author{Yaron Amouyal}\affiliation{\HaifaMat}
\author{Lukasz Plucinski}\affiliation{\FZJ}
\author{Efrat Lifshitz}\affiliation{\Haifa} 
\author{Krzysztof Wohlfeld}\affiliation{\IFT}
\author{Magdalena Birowska}\affiliation{\IFT}
\author{Markus Morgenstern$^*$}\affiliation{\AC}
\date{\today} 

\def\MMM{\textcolor{red}}
\def\MM{\textcolor{blue}}
\def\BB{\textcolor{blue}}
\def\Jeff{\textcolor{pink}}
\def\Magda{\textcolor{violet}}
\def\Milosz{\textcolor{green}}


\renewcommand{\textcolor}[2]{#2}


\begin{abstract}
\noindent \textbf{Abstract}\\
NiPS$_3$ is an exfoliable van-der-Waals intralayer antiferromagnet with zigzag-type spin arrangement. It is distinct from other TMPS$_3$ (TM: transition metal) materials 
by
optical excitations into a strongly correlated state that is tied to the magnetic properties. 
However, the related, fundamental band structure across the antiferromagnetic phase transition has not been probed yet. 
Here, we use angular-resolved photoelectron spectroscopy with $\mu$m resolution in combination with DFT+U calculations 
for that purpose. 
We identify a characteristic band shift 
across $T_{\rm N}$. It is attributed to bands of mixed Ni and S character related to the superexchange interaction of Ni 3t$_{\rm 2g}$ orbitals. 
Moreover, we find a structure above the valence band maximum with little angular dispersion that could not be reproduced by the calculations. The discrepancy suggests the influence of many-body interactions beyond the DFT+U approximations in striking contrast to the results on MnPS$_3$ and FePS$_3$, where these calculations were sufficient for an adequate description. 

\end{abstract}
 

\maketitle
\noindent {{$^\dag$}These authors contributed equally to this work.
{$^*$}Corresponding author: M.~Morgenstern, Email: \href{mmorgens@physik.rwth-aachen.de}{mmorgens@physik.rwth-aachen.de}} 

\section{Introduction}

%
\begin{figure*}[!htb]
\centering
\includegraphics[width=\textwidth]{Fig_1.pdf}
\vspace{-0.5 cm}
\caption{{\bf Characterization of exfoliated NiPS$_3$ flakes.}
(a) Atomic top view (top image) and side view (middle image) of 
two layers of NiPS$_3$, red arrows: spins of the Ni atoms, bottom image: only the Ni atoms highlighting their honeycomb atomic and zigzag-type AFM magnetic arrangement. 
(graphics made by VESTA \cite{VESTA}). (b) Optical microscope image of an exfoliated NiPS$_3$ on a Au/Ti/SiO$_2$/Si substrate. The yellow dashed circle marks the 15 layer thick area used for ARPES. (c) Atomic force microscopy image of the flake area probed by ARPES (rms roughness: 0.14\,nm). (d) XPS of the same 15 layer area after the ARPES measurements, \textit{h}$\nu$ = 220\,eV, peaks  are labeled by element and atomic core level orbital. Inset: XPS at \textit{h}$\nu$ = 230 eV exposing the Ni 3p peak by shifting the otherwise overlapping S-Auger peak to lower $E-E_{\rm F}$. (e) Raman spectra at room temperature (RT) and 77\,K ($T_{\rm N}=155$\,K) labeled according to their symmetries and as P1 to P9 for easier description. A Si peak originating from the substrate is also labeled \cite{Kuo2016}. The peak around 730 cm$^{-1}$ is due to a second-order Raman mode \cite{Kuo2016}. 
Insets: Zoom into the area of P2 
with apparent peak splitting at 77 K caused by magnon-phonon interaction \cite{Muhammad2023,Kim2019b}. 
}
\label{Fig. 1}
\end{figure*}
Van der Waals (vdW)-type two-dimensional (2D)  ferromagnets (FMs) and antiferromagnets (AFMs) 
\cite{Huang2017,Sun2019b,Gong2017,Wang2018d,Deng2018,Fei2018} enrich 
the possibilities for spintronics, magneto-optical devices and magnetic sensors \cite{Li2019,Zhang2019,Burch2018,Gong2019, Huang2020}. This is related to the fact that
the 2D magnetism can be tuned 
via magnetic fields, electrostatic gating, current pulses, strain, light, ion intercalation, proximity and moiré lattices 
\cite{Burch2018,Li2019,Xu2021,Mi2022,Shin2022,Ahn2024,Cham2024} as well as due to the relatively strong magneto-electric, magneto-elastic and magneto-optic couplings  \cite{Wang2022,Vaclavkova2021, Ressouche2010, Kang2020,Hwangbo2021,Dirnberger2022}. 
Moreover, these materials enable a novel access to fundamental questions in 2D magnetism such as
the Berezinskii–Kosterlitz–Thouless (BKT) transition or proximity-induced modifications of the magnetization 
\cite{Gong2017,Burch2018,Gong2019,Huang2020,Srivastava2020,BedoyaPinto2021,Albarakati2022,Ahn2024,Sun2024}. 

A unique type are  transition metal (TM) phosphorus trisulfides TMPS$_3$ 
\cite{Joy1992,Wang2018c}. In contrast to other vdW magnets, they are intralayer AFMs and, moreover, rather directly realize the three most prominent classes of 2D magnetism (Ising, XY, Heisenberg) \cite{Chittari2016,Kim2019b,Autieri2022}. Hence, the materials provide a versatile platform to study 2D AFM properties. The TMPS$_3$ materials  are  semiconductors with a layered honeycomb structure of the TM$^{2+}$ atoms, each surrounded by covalently bonded (P$_2$S$_6$)$^{4-}$ bipyramids (Fig.~\ref{Fig. 1}a). 
The magnetism arises from the competition between direct exchange interactions between neighboring TM atoms and indirect superexchange  mediated by S and P atoms \cite{Autieri2022,Yan2023,Rybak2024}. Depending on the relative strength of the competitive interactions, 
various long-range AFM orders have been found, dubbed zigzag, Néel, and stripy AFM order \cite{Joy1992,Sivadas2015}.

NiPS$_3$ stands out since it exhibits significant spin-charge coupling, leading to spin-correlated excitons whose origin is actively debated. The excitons display layer-dependent characteristics and polarization properties correlated with the material's magnetism \cite{Kim2018,Kang2020,Lane2020,Wang2021,Hwangbo2021,Wang2022,Jana2023,Klaproth2023,Song2024,Wang2024,Allington2025}. \MM{This might enable novel optomagnonic interconnects (Supplementary section S1), e.\,g. via the ultra-sharp exciton peak observed at approximately 1.47\,eV \cite{Chittari2016,Kim2018,Muhammad2023,Du2016}.}
\MM{It has also been shown that the properties of this exciton peak can only be described correctly by calculations employing muliplet-type exact diagonalization 
\cite{Klaproth2023}.}
Moreover, the magnetic interactions are surprisingly dominated by a third nearest neighbor AFM exchange coupling  
due to a substantial overlap of p- and d-orbitals along the corresponding Ni-S-P-S-Ni interaction path \cite{Scheie2023,Autieri2022,Chittari2016,Yan2023}. This leads to an intralayer AFM zigzag order as the ground state below the bulk Néel temperature  $T_\text{N}$ $\approx$ 155 K, where the spins are oriented mostly in-plane along the zigzag direction (Fig.~\ref{Fig. 1}a) with a slight canting out-of-plane by $\sim$8$^{\circ}$ \cite{Wang2022}. 
In monolayer form, the long-range magnetic order
is most likely suppressed potentially leading to a BKT transition \cite{Kim2019b,Sun2024}. This behavior aligns with descriptions based on a highly anisotropic XXZ model \cite{Joy1992,Wildes2015}.

\begin{table*}[t]
    \centering
    \resizebox{0.8\textwidth}{!}{%
    \begin{tabular}{|c|c|c|c|c|c|}  
        \hline
        \textbf{Material} & \textbf{AFM \newline order} & \textbf{Bandgap\ (eV)} & \textbf{$T_{\rm N}$\,(K)} & \textbf{$U_{\rm eff}$\,(eV)} & \textbf{Band shift across $T_{\rm N}$} \\
        \hline
        NiPS$_3$\newline(this work) & zigzag\newline(XY-like) & 1.8 & 155 & 1.6 & mixed Ni~3d/S~3p band, extra\\
        &&&&& band at valence band maximum\\
        \hline
        FePS$_3$~\cite{Pestka2024} & zigzag\newline(Ising-like) & 1.5 & 117 & 1.2 & S~3p-, Fe~3d- and P~3p-type bands \\
        \hline
        MnPS$_3$~\cite{Strasdas2023} & Néel & 2.94 & 78 & 1.8 & Mn~3d dominated band\\
        \hline
    \end{tabular}}
    \vspace{0mm}
    \caption{\MM{TMPS$_3$ properties, $T_{\rm N}$: Néel temperature, $U_{\rm eff}$: selected effective Hubbard parameter of the transition metal d levels for the DFT+U calculations that  reproduces the experimental $\mu$-ARPES data most favorably, band shifts across $T_{\rm N}$  are determined by $\mu$-ARPES and assigned to elements and orbitals by comparison with the band structures from DFT+U calculations.}}
    \label{tab:tab1}
\end{table*}

The magnetic ordering of NiPS$_3$ has already been probed by multiple methods  \cite{Kim2018,Kim2019b,Kang2020,Hwangbo2021,Afanasiev2021,Basnet2021,Basnet2022,Scheie2023,Muhammad2023}, but none of them is sensitive to details of the band structure, which has only been probed at room temperature so far, i.\,e. above $T_{\rm N}$, using angular resolved photoelectron spectroscopy (ARPES) \cite{Yan2023b,Klaproth2023,Nitschke2025}.
Here, we provide band structure mapping by ARPES at variable temperature $T$ above and below $T_{\rm N}$ leveraging the approach that we recently developed for the isostructural MnPS$_3$ and FePS$_3$ \cite{Strasdas2023,Pestka2024}. We prevent sample charging due to photoelectrons, that often appears for semiconductors at low $T$ \cite{Bianchi2023,DeVita2022, Nitschke2023,Koitzsch2023, Voloshina2023}, by exfoliating thin films of NiPS$_3$ onto a conductive Au film deposited on Si/SiO$_2$, itself necessary for identifying the flakes optically. 
The resulting band structure is compared with
density functional theory calculations (DFT+U) enabling the identification of the elemental and orbital character of the different bands and revealing an optimal Hubbard parameter $U_{\rm eff}=1.6$\,eV. 
Most importantly, we observe a pronounced band shift across $T_\text{N}$ for a valence band 
identified as a mixed Ni 3d/S 3p band that could be attributed to a next nearest neighbor superexchange path. 
Moreover, we reveal an additional band feature in ARPES that appears above the valence band maximum, but could not be reproduced by the DFT+U calculations. We believe that it testifies the strong correlations in NiPS$_3$ compared to MnPS$_3$ and FePS$_3$ in line with the optical results \cite{Lane2020,Wang2022,Kim2018,Kang2020,Hwangbo2021}. 

\MM{Since the comparison between these three materials is crucial for the missing assignment of this band, we provide such a comparison in table~\ref{tab:tab1}. In particular, the selected $U_{\rm eff}$ values are rather similar for all three materials as expected by the identical chemical environment of the transition metal 3d levels.} 



\section{Results and Discussion}

\begin{figure*}[!thbp]
\centering
\includegraphics[width=\textwidth]{Fig_2.pdf}
\vspace{-0.3 cm}
\caption{{\bf Electronic band structure change of NiPS$_3$ across \textit{T}$_\text{N}$.}
(a), (b) Raw ARPES data  above (220\,K) and below (45\,K) \textit{T}$_\text{N}=155$\,K,  white arrows point to the changes 
highlighted in c--h, \textit{h}$\nu$ = 60\,eV, $\overline{\rm M}\overline{\Gamma}\overline{\rm M}$ direction. (c) Intensity plots from a--b at $k_\parallel=0/$\AA\ after smoothing by box filters ($100$\,meV/two pixels 
in $E$, $0.041/$\AA/5 pixels in $k_\parallel$). An additional peak appears at $E-E_{\rm F}=-3.7$\,eV close to the peak at $-4.15$\,eV, which is present at both $T$ (arrows). 
(d), (e) Smoothed ARPES data from a--b after subtracting a 3$^{\rm rd}$-order polynomial background, adapted  for each $k_\parallel$ 
to the whole displayed $E-E_{\rm F}$ range as shown exemplarily in f.  White arrows (same energy as in a--b)  mark the most prominent change. 
(f) Zoom into c (blue and red curve) with a 3$^{rd}$ order polynomial fit to the blue curve (dashed line). 
(g), (h) Angular averaged intensity of the smoothed data 
in the energy range marked by a box in c, (g) above \textit{T}$_\text{N}$, (h) below \textit{T}$_\text{N}$, \MM{black, pink arrows: same energies as in c, green arrow: energy of intensity at 220\,K that likely shifts upwards at 45\,K.}  
Insets: (\textit{k}$_\textit{x}$, \textit{k}$_\textit{y}$) plots of the smoothed photoelectron intensity at $E-E_{\rm F}=-3.7$\, eV (green, pink arrows in main image) after background subtraction. The center is $\Gamma$. (a)-(c), respectively (d)-(f), share the same energy axis. 
}
\label{Fig. 2}
\end{figure*}

NiPS$_3$ single crystals are synthesized by the vapor transport method.
Flakes are subsequently exfoliated by the scotch tape method onto Au/Ti covered Si/SiO$_2$ at 60$^\circ$\,C, shortly after the substrate was cleaned by plasma ashing \cite{Strasdas2023} (Materials and Methods). Flakes with appropriate thickness and lateral size are identified using optical and atomic force microscopy (Fig.~\ref{Fig. 1}b-c). 
Typically, these flakes exhibit various thicknesses being most thin at the rim.
In this paper, we studied a 15 layer thick area 
with a lateral size of $17 \times 21\,\mu$m$^2$ as marked in Fig.~\ref{Fig. 1}b. Atomic force microscopy reveals a rms roughness  of $0.14$\,nm and no obvious contaminations in that area (Fig.~\ref{Fig. 1}c). The relatively small thickness is crucial to avoid charging during $\mu$-ARPES at low temperature as detected, e.\,g. for 45 layers at 90\,K, where all bands are shifted downwards by 0.3\,eV. 

Figure~\ref{Fig. 1}d displays  X-ray photoelectron spectroscopy (XPS) data of the 15 layer area. 
All components of NiPS$_3$ appear as core level peaks \cite{Yan2021,Curr1995,Piacentini1982} as well as a S Auger peak \cite{Coad1972}. The latter shifts with respect to the other peaks by changing the photon energy $h\nu$ exposing the Ni 3p peak \cite{Gorham2012}. A small contribution from the spin-split Au 4f peaks \cite{Gorham2012} is also visible likely originating from the substrate due to the relatively large spot of the photon beam on the sample (diameter: 15\,$\mu$m) at the large $h\nu$ for XPS. 


\begin{figure*}[!thbp]
\centering
\includegraphics[width=1\textwidth]{Fig_3.pdf}
\vspace{-0.8 cm}
\caption{{\bf Identifying the orbital character of the bands.}
 (a) Surface projected Brillouin zone of the atomic geometry (grey hexagon) with high symmetry points marked and the magnetic zigzag structure (blue rectangle).
(b) Band structure of NiPS$_3$ in the AFM zigzag configuration as sketched on top, DFT+U calculations unfolded to the hexagonal, atomic Brillouin zone sketched in a \cite{Popescu2012}, $U_{\rm eff} = 1.6$\,eV, \textit{k}$_\textit{z}$ = 0.1/\AA $,\overline{\rm M}\overline{\Gamma}\overline{\rm M}$ direction as marked in a. 
\MM{The energy is related to the valence band maximum $E_{\rm VBM}$.} The atomic orbital contributions of each state are depicted by overlapping, colored circles with diameter proportional to the contribution (color code on the left, diameter of dark green circles at \MM{$E-E_{\rm VBM}\approx-5$\,eV} are $\sim 100$\,\%.  
(c) Same as (b) but only showing the s, p$_z$ and d$_{z^2}$ orbital contributions in accordance with simplified ARPES selection rules \cite{Moser2017}. Blue arrow marks the position of band change observed in Fig.~\ref{Fig. 2} (d) ARPES curvature plot, \textit{h}$\nu$ = 60 eV, $T = 45$\,K, $\overline{\rm M}\overline{\Gamma}\overline{\rm M}$ direction. Labels i-ix: similar band features in c (DFT+U) and d (ARPES) (see text). Dashed rectangles: area of the magnetically induced band shift  (Fig.~\ref{Fig. 2}).
(e) Same as c for the Néel configuration as sketched on top.
%
Figures b--e share \MM{the same scaling of the energy}. The DFT + U data are adapted to the ARPES data by a rigid shift optimizing the overlap of structures i-ix. 
}
\label{Fig_3}
\end{figure*}

The antiferromagnetic phase transition of the exfoliated NiPS$_3$ flakes is verified by Raman spectroscopy 
(Fig.~\ref{Fig. 1}e). 
Due to the nearly realized D$_{\rm 3d}$ intralayer point group symmetry \cite{Kuo2016}, NiPS$_3$ has eight major Raman-active phonon modes, 
all visible in our Raman spectra with frequencies in good agreement with the literature \cite{Kim2019b,Wang2019,Muhammad2023,Kuo2016}. 
We reproduce three signatures that have previously been assigned to the AFM phase transition \cite{Kim2019b, Muhammad2023}. Firstly, a broad 2-magnon scattering peak, centered at $\sim$530 cm$^{-1}$, appears below $T_{\rm N}$ at 77 K and is absent at room temperature (RT) \cite{Kim2019b}. Secondly, the peak P2 caused by in-plane vibrations of the Ni atoms splits into a double peak (insets, Fig.~\ref{Fig. 1}e) as attributed to a pronounced magnon-phonon coupling  \cite{Kim2019b,Muhammad2023}. Thirdly, the peak P8, 
caused by an in-plane E$_{\rm g}$-type vibration of the (P$_2$S$_6$)$^{4-}$ bipyramids, develops a Fano-resonance line shape below \textit{T}$_\text{N}$ due to an interference between the phonon excitation and the continuum of 2-magnon excitations \cite{Fano1961}. These three fingerprints confirm the AFM phase transition in our NiPS$_3$ flakes.  


Figure~\ref{Fig. 2} shows the main result of this work, namely a band shift across \textit{T}$_\text{N}$ near the $\overline{\Gamma}$ point. We present the raw ARPES data (a--b), the data after box-type smoothing (c, only the energy distribution curve (EDC) is shown), after additional background subtraction (d--f) and angularly averaged in $\boldsymbol{k}$-space around the $\overline{\Gamma}$ point (g--h). Already the raw data (Fig.~\ref{Fig. 2}a--b) exhibit a slightly enhanced intensity below $T_{\rm N}$ in between the white arrows. This change can be identified as an additional peak in the EDC at $\overline{\Gamma}$ after mild smoothing (upper arrow, Fig.~\ref{Fig. 2}c) implying a state at $E-E_{\rm F}=-3.7$\,eV  below $T_{\rm N}$ ($E$: energy, $E_{\rm F}$: Fermi energy as probed on Au). A polynomial background subtraction (Fig.~\ref{Fig. 2}f) increases the visibility of that peak. The resulting energy vs. wave vector display of the photoelectron intensity (Fig.~\ref{Fig. 2}d--e) reveals that the peak belongs to a rather flat band extending up to an in-plane wave vector $|\boldsymbol{k}_\parallel|\approx 0.15/$\AA, where it merges with upwards dispersing bands. These features are corroborated by the angulary averaged data in $\boldsymbol{k}_\parallel$ space  (Fig.~\ref{Fig. 2}g--h). The  band at $E-E_{\rm F}=-3.7$\,eV  is again visible below $T_{\rm N}$ (Fig.~\ref{Fig. 2}h, pink arrow). \MM{It shifts downwards by roughly 150\,meV above $T_{\rm N}$ (Fig.~\ref{Fig. 2}g, green arrow), nearly merging with the band at $E-E_{\rm F}=-4.15$\,eV (Fig.~\ref{Fig. 2}g-h, black arrows).}  The insets finally show
the two-dimensional $\boldsymbol{k}_{\parallel}$ plots at that energy exhibiting a ring above $T_{\rm N}$ due to the upwards dispersing bands  and a full circle below $T_{\rm N}$ due to the additional flat band. The upwards dispersing bands visible in the main image are also slightly less steep below $T_{\rm N}$.

\MM{One might suspect that the downwards shift is related to a distinct structural phase transition or to a continuous thermal contraction of the NiPS$_3$ crystal. However, there is no structural phase transition in that temperature range as known from neutron diffraction \cite{Wildes2015}, while shifts of electronic states by thermal contraction are less than 50\,meV between 220\,K and 45\,K as deduced from electronic Raman spectroscopy \cite{Wang2022}}. \MMM{We confirmed this by DFT+U calculations using the lattice parameters deduced from neutron diffraction at 295\,K and 2\,K to scale the lattice parameters in DFT+U. These calculations did not reveal any band shifts larger than 50\,meV as well (not shown), in particular in the energy range where the band structure change was observed in Fig.~\ref{Fig. 2}.}   

Next, we compare the ARPES data with DFT+U calculations of the magnetic ground state (AFM zigzag type) and other magnetic configurations
(AFM Néel and disordered). This aims to identify the elemental and orbital character of the different bands, in particular the ones that are changing across $T_{\rm N}$. We started by  selecting the proper Hubbard $U_{\rm eff}$ parameter (DFT+U)  by detailed comparison with the experimental data.
The best match for the ARPES data at $h\nu =60$\,eV and $T=45$\,K is found for $U_{\rm eff}=1.6$\,eV (Suppl. section S3) and $k_z=0.1/$\AA\, (Suppl. section S4). 
\MM{The $k_z$ selection is corroborated by photon energy dependent ARPES data that are successfully compared with the $k_z$ dependent DFT+U data employing an inner potential of 12.1\,eV (Suppl. Fig. S5).}

%

\begin{figure*}[!hbtp]
\centering
\includegraphics[width=0.9\textwidth]{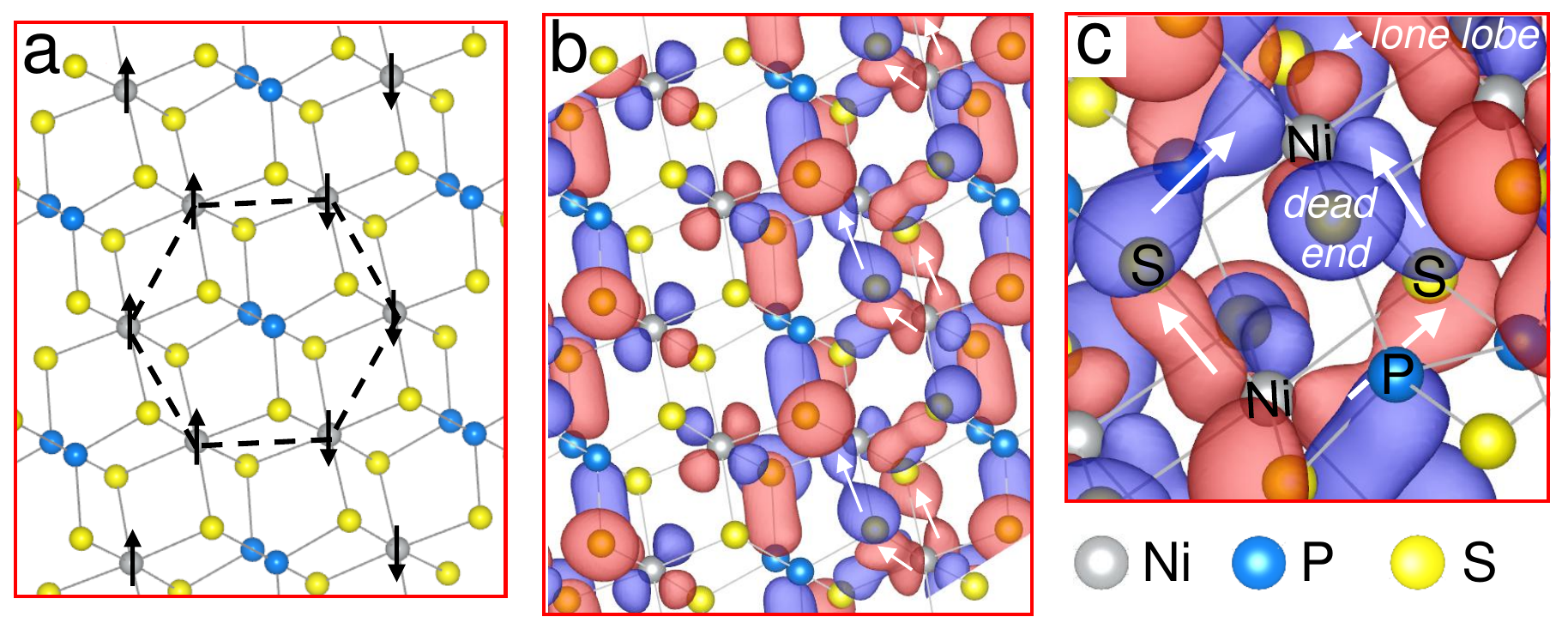}
\vspace{-0.3 cm}
\caption{\textbf{Charge density of magnetically shifted band.} (a) Crystal structure of a single NiPS$_3$ layer viewed along the c axis (Fig.~\ref{Fig. 1}a)  with a marked honeycomb of Ni atoms (dashed line) and spin directions in the zigzag AFM configuration (black arrows). (b) Contour planes of the charge density $|\psi(\boldsymbol{x})|^2$ of a state at $E-E_{\rm F}=-3.9$\,eV and $\boldsymbol{k}=\boldsymbol{0}$/\AA\, (marked by blue arrow in Fig.~\ref{Fig_3}c). This state likely represents the state that is shifting across $T_{\rm N}$. The area and angle of view are the same as in a. Blue and red contour planes mark opposite signs of the wave function. The white arrows highlight the two superexchange paths along Ni 3t$_{\rm 2g}$-S 3p-S 3p-Ni 3t$_{\rm 2g}$ between two ferromagnetically coupled Ni atoms. (c) Zoom into b at a tilted angle of view showing the two supexchange paths more clearly (arrows) as well as the lone lobe of the Ni 3t$_{\rm 2g}$ orbital and the remaining Ni 3t$_{\rm 2g}$ lobe connected to a S 3p-type dead end orbital (see Suppl. Video). (graphics made by VESTA \cite{VESTA}).
}
\label{Fig_3b}
\end{figure*}

The band structure is additionally unfolded to the atomic Brillouin zone (grey hexagon in Fig.~\ref{Fig_3}a) \cite{Popescu2012} (Suppl. Fig.~S2). This reflects the fact that the enlarged periodicity by AFM order is barely imprinted in the wave functions  \cite{Hüfner2013}. 
Indeed, $(k_x,k_y)$ maps of the ARPES data did not show any patterns reminiscent of the magnetic Brillouin zone, but only of the hexagonal, geometric one (Suppl. section S2). Figure~\ref{Fig_3}c shows the same DFT+U data as in b, but after selecting only the orbital contributions s, p$_z$, and d$_{z^2}$ within a frame where the $z$-direction is pointing perpendicular to the layers. According to simplified selection rules assuming plane waves as final states \cite{Moser2017}, these orbital contributions are excited preferably for our light polarization geometry \cite{Strasdas2023}. \MM{Orbital contributions to the different bands beyond the simplified selection rules are analyzed in Suppl. section S5 with the results partially included in the following discussion.} Figure~\ref{Fig_3}e shows the same selection of bands for the Nèel configuration \MM{which is not realized in NiPS$_3$. This additional band structure serves to identify bands that are changing with the magnetic arrangement of the atoms. For the same purpose, we also provide the calculated band structure of a disordered magnetic configuration in Suppl. Fig.~S11. Finally,}  Fig.~\ref{Fig_3}d shows the ARPES curvature data below $T_{\rm N}$. 
%
 The curvature is used for better visibility of the various features (Suppl.\ section\ S12). We always cross-check with the raw data as in Fig.\ \ref{Fig. 2} to corroborate that a feature is not an artifact of the curvature derivation. 

 Multiple features are similar between the ARPES results (Fig.~\ref{Fig_3}d) and the DFT+U data \MM{of the zigzag configuration} (Fig.~\ref{Fig_3}c) as labeled by i-ix. 
 The bright feature (i) and the nearly parabolic feature (ix) are used to align the Fermi level of the ARPES data to the DFT+U band structure. The additional structure in ARPES at $E-E_{\rm F}\approx -1.3$\,eV is discussed below. The band (i) has a dominating S 3p character with contributions from Ni 3d$_{xz,\,yz,\,xy,\,x^2-y^2}$ (Suppl. Fig. S6). It  
 is rather flat in calculation and experiment.   
The  lowest energy band (ix) has a parabolic dispersion close to $\overline{\Gamma}$ in both cases and a mixed P 3p$_z$ and S 3p character. This band appears quite similarly for MnPS$_3$ and FePS$_3$ \cite{Strasdas2023,Pestka2024}. It is slightly lower in the DFT+U data than in the ARPES data. This is probably due to the facts that firstly $k_{z}$ moves upwards with decreasing $E-E_{\rm F}$ in ARPES at constant $h\nu$ and secondly this band moves upwards with increasing $k_{z}$ in the DFT+U data (Suppl. Figs. S4/S5). At slightly larger $E-E_{\rm F}$ are bands with nearly exclusive S 3p$_{x,y}$ character (viii)  that are barely visible in ARPES in accordance with the simplified selection rules \cite{Moser2017}. The remaining features ii-vii exhibit a mixed Ni 3d/S 3p character. In ARPES, they are dominated by a large cross-type structure gapped at $\overline{\Gamma}$ (iv,vi) with a rather flat band on top (ii). A flat band of strong Ni 3d$_{z^2}$ character (ii) and a gapped crossing of mixed character (iv,vi) are also visible in DFT+U.
Other features are more subtle and partially distinct such as the relatively steep lower part of the cross from vi to vii in ARPES which is not continuous across the whole energy range in DFT+U. Nevertheless, the agreement in the whole energy range below $E-E_{\rm F}=-1.5$\,eV is rather satisfactory regarding the fact that the band structure of the calculation was only rigidly shifted for adaption. 

This importantly implies that the magnetically induced band shift identified in Fig.~\ref{Fig. 2} is at the gapped crossing of feature (iv) and (vi), more precisely it is the upper part (iv) at $E-E_{\rm F}=-3.7$\,eV around $\overline{\Gamma}$ in the ARPES data
(Fig.~\ref{Fig_3}d). \MM{Accordingly, it is at $E-E_{\rm VBM}=-2.2$\,eV in the DFT+U data (arrow in Fig.~\ref{Fig_3}c)}.
Consistently, these band structure features exhibit a relatively strong change in DFT+U data, if the magnetic structure is changed. \MM{For the zigzag configuration, the gapped area at $\overline{\Gamma}$ is $\sim 0.3$\,eV wide (arrow in Fig.~\ref{Fig_3}c), while it is increased to $\sim 0.4$\,eV for the Néel configuration (Fig.~\ref{Fig_3}e, within dotted box).} For a disordered magnetic structure, we find that  features (iv,vi) get shifted upwards by $\sim 100$\,meV with respect to the zigzag configuration (Suppl. Fig.~S11). 
\MM{Data along the $\overline{\rm K}\overline{\Gamma}\overline{\rm K}$ direction are very similar to the  $\overline{\rm M}\overline{\Gamma}\overline{\rm M}$ direction (Suppl. Fig.~S9)  and, hence, support all our conclusions.}

The four bands in the energy region of the observed band shift consist of all Ni 3d and S 3p orbitals as well as of a minor contribution from P 3p$_{z}$  (Suppl. section S5). However, only one of these bands features a bonding configuration between neighboring Ni atoms. Figure~\ref{Fig_3b}b shows the corresponding charge density for the state at $\overline{\Gamma}$ (charge density of the other three bands, Suppl. section S6). Using the octahedral coordinate system with $z$ along a Ni-S bond, one observes Ni 3t$_{\rm 2g}$ orbitals (d$_{xy}$, d$_{xz}$, d$_{yz}$) for both Ni zigzag chains, distinct by their spin orientation. On the left chain with spin $\uparrow$, the Ni 3t$_{\rm 2g}$ orbitals do not have any overlap with neighboring orbitals. In contrast, for the right chain, three lobes of each Ni 3t$_{\rm 2g}$ orbital overlap with a lobe of an adjacent S 3p orbital. The zoom  in Fig.~\ref{Fig_3b}c as well as rotated views (Suppl. movie)  reveal that two of these S 3p orbitals provide a connection path between two Ni atoms (arrows in Fig.~\ref{Fig_3b}b), each via the two lobes of a single S 3p orbital. The third connected S 3p orbital (lower lobe of the upper Ni atom  in Fig.~\ref{Fig_3b}c) is, however, a dead end, i.\,e. the S 3p orbital does not have any connection to other atoms (Suppl. movie). The final forth lobe of the  Ni 3t$_{\rm 2g}$ orbital is not connected at all and, hence, called a lone lobe as marked in Fig.~\ref{Fig_3b}c.

%
The superexchange path between two Ni atoms by only one S orbital on both sides implies that the interaction is antiferromagnetic. This naturally explains that the band is shifted upwards in energy when these chains become ferromagnetic within the zigzag configuration making the particular bond more unfavorable. It remains unclear why we observe only this change of a band and none of the spin-polarized 3e$_{\rm g}$ levels at higher energy. We conjecture that this is related to the selection rules in our ARPES geometry not being sensitive to the corresponding levels (Suppl. Fig.~S6). 

\begin{figure*}[!hbtp]
\centering
\includegraphics[width=1\textwidth]{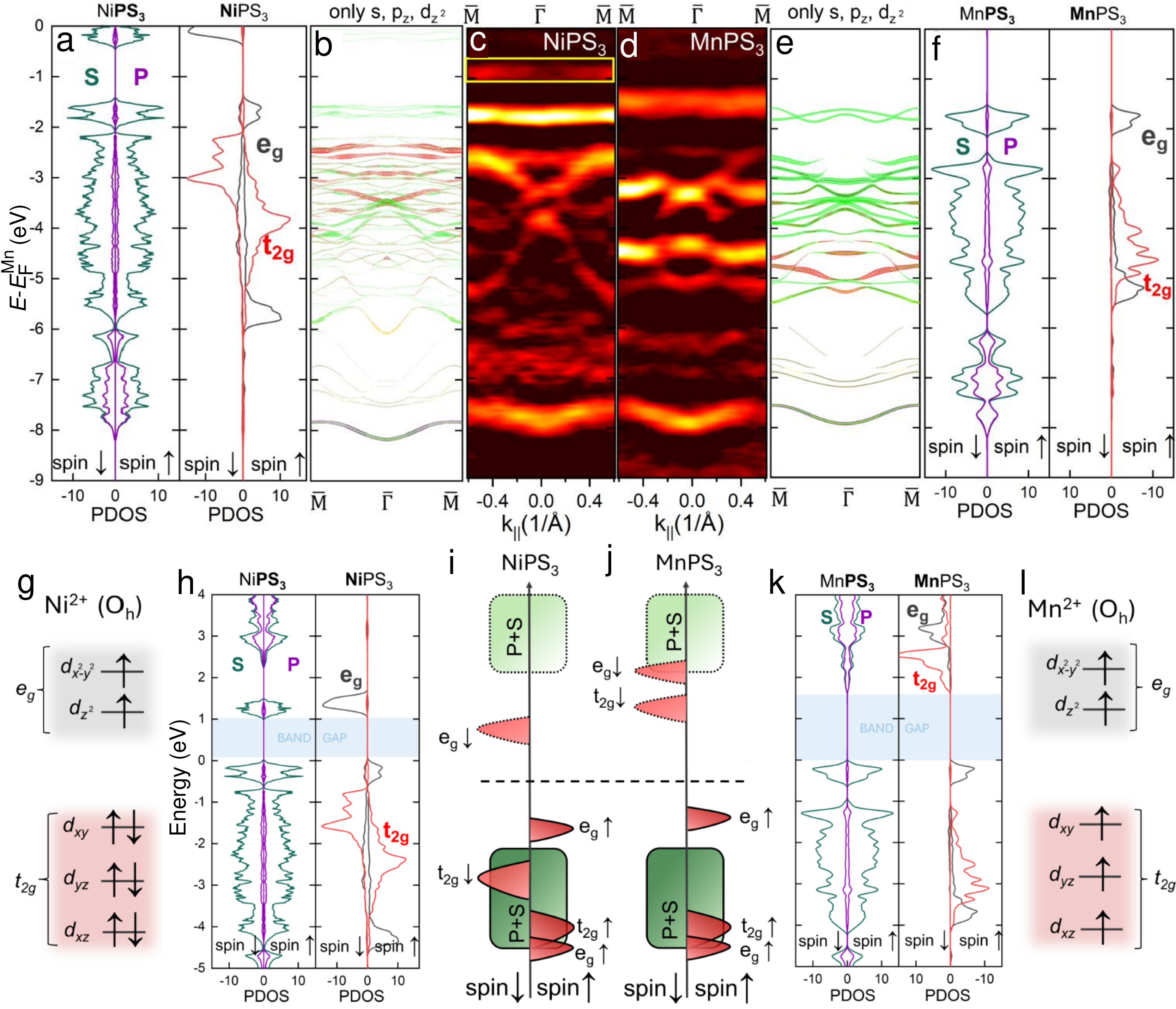}
\vspace{-0.3 cm}
\caption{\textbf{Comparing NiPS$_3$ and MnPS$_3$ \cite{Strasdas2023}.} (a) Spin resolved partial density of states (PDOS) for NiPS$_3$, left: ligand atoms S, P, right:  Ni 3d orbitals t$_{\rm 2g}$ (d$_{xy}$, d$_{xz}$, d$_{yz}$), $e_{\rm g}$ (d$_{x^2-y^2}$, d$_{z^2}$), octahedral coordinate system (see text), DFT+U, $U_{\rm eff}=1.6$\,eV, $k_z=0.1/$\AA. (b) Band structure corresponding to a after unfolding to the atomic, hexagonal Brillouin zone and selecting the s, p$_z$ and d$_{z^2}$ contributions, $z$ along layer normal (same as Fig.~\ref{Fig_3}c) (c) ARPES curvature of NiPS$_3$, $h\nu=60$\,eV, $T=45$\,K (same as Fig.~\ref{Fig_3}d, but shifted upwards by 0.4\,eV to match better to MnPS$_3$) 
(d) same as c for MnPS$_3$ without energy shift, $h\nu=50$\,eV, $T=43$\,K ($T_{\rm N}=78$\,K) \cite{Strasdas2023}. (e) same as b for MnPS$_3$, $U_{\rm eff}=1.8$\,eV, $k_z=0.46/$\AA\, \cite{Strasdas2023}. (f) Same as a for MnPS$_3$. (g) Schematic occupation of Ni$^{2+}$ 3d levels in NiPS$_3$. (a)-(c) are adapted to the Fermi level $E_{\rm F}^{\rm Mn}$ of MnPS$_3$ by a rigid energy shift.  (h) Same as a including the conduction band states. (i) Simplified spin polarized PDOS of NiPS$_3$. (j)-(l) Same as g-i for MnPS$_3$ in reverse order.
}
\label{Fig_4}
\end{figure*}
We note in passing that we did not find any AFM induced \textit{C$_3$} symmetry breaking in the band structure of NiPS$_3$ by  ARPES. The three different  $\overline{\Gamma}\overline{\rm M}$ or $\overline{\Gamma}\overline{\rm K}$ directions are identical within error bars, albeit the antiferromagnetic zigzag structure breaks the \textit{C$_3$}  symmetry (Fig.~\ref{Fig. 1}a). Since the DFT+U calculations do not reveal any symmetry breaking \MM{that is significantly larger than the energy resolution of the experiment (50\,meV), the effect is likely too weak to be detected (Suppl. section S8)}. In ARPES, we can, moreover, not exclude that the relatively small single domains (1-10\,$\mu$m) \cite{Lee2024} partially overlap within the extraction spot of the photoelectrons (diameter: 5\,$\mu$m)
\cite{He2024}. 


%

This reasonable interpretation leaves us with the question of the origin of the flat feature in ARPES at $E-E_{\rm F}=-1.3$\,eV, that is not captured by DFT+U. It appears as a shoulder in the raw ARPES data at energies above the valence band maximum and is nearly identical above and below $T_{\rm N}$ (Suppl. Fig~S12). It, moreover, exhibits the identical photon energy ($k_z$) dependence as the valence band maximum (Suppl. Fig.~S5, S12).
To substantiate its discussion, we compare with the data of MnPS$_3$ \cite{Strasdas2023} (Fig.~\ref{Fig_4}), where all observed features in the ARPES data could be identified with bands of the DFT+U calculation.
Figure~\ref{Fig_4}a-f displays the direct comparison using DFT+U data within the respective antiferromagnetic ground state. The figure includes the partial density of states (PDOS) (Fig.~\ref{Fig_4}a,f) where the DOS is projected to the S and P atoms (left) or the e$_{\rm g}$ (3d$_{z^2,\,x^2-y^2}$) and t$_{\rm 2g}$ (3d$_{xy,\,xz,\,yz}$) orbitals of Ni and Mn (right), respectively, using the octahedral coordinates with $z$ along a Ni-S bond. 
Moreover, the PDOS of minority and majority spins is plotted separately. Finally, the ARPES and DFT+U data of NiPS$_3$ are shifted upwards by 0.4\,eV with respect to their $E-E_{\rm F}$ to align the low energy bands of the two materials. Hence, $E_{\rm F}^{\rm Mn}$, the Fermi level of MnPS$_3$, is chosen as reference for for the energy scale.

We find large similarities between NiPS$_3$ and MnPS$_3$. For example, the upper valence band of S 3p/Ni 3d character, respectively S 3p/Mn 3d character is nearly identical in strength of dispersion and orbital contributions.
Moreover, the lowest energy bands ($E-E_{\rm F}^{\rm Mn}<-7.5$\,eV) of mixed S 3p/P 3p character are very similar including the relative contribution of S and P orbitals. The energy range in between is, however, different. Firstly, there is a full spin polarization of the t$_{\rm 2g}$ orbitals for MnPS$_3$ that is absent for NiPS$_3$. Related, the mixed 3d$_{z^2}$/S 3p bands are more confined in energy for MnPS$_3$. In particular, the two bands dominated by Mn 3d$_{z^2}$ orbitals ($E-E_{\rm F}^{\rm Mn}\in[-5,-4]$\,eV ) are the ones that are most strongly changed across $T_{\rm N}$ \cite{Strasdas2023}. In contrast, for NiPS$_3$, all bands with $E-E_{\rm F}^{\rm Mn}\in[-5.5,-2]$\,eV  have strong, often dominating contributions from Ni 3d orbitals (Suppl. section S5). This difference is in line with the 3d level occupation (see below).
Most importantly, the experimental data is very reasonably fitted by DFT+U for both data sets except for the barely dispersing feature at $E-E_{\rm F}^{\rm Mn}\approx -1$\,eV for NiPS$_3$ (highlighted by the yellow box in Fig.~\ref{Fig_4}c). The correspondence between ARPES and DFT+U is also reasonable for FePS$_3$, albeit a stretch of DFT+U data by 7\,\% was required for a quantitative match \cite{Pestka2024}.

\MM{To corroborate that the extra feature (yellow box, Fig.~\ref{Fig_4}c) is not captured by DFT-type calculations, we also performed calculation with other approaches to treat the exchange-correlation energies. We employed HSE while varying the percentage of the Hartree-Fock exchange contribution as well as G$_0$W$_0$ and EVGW$_0$ based on a starting PBE band structure (Suppl. section S13). The results feature various shifts of the Ni 3d PDOS, but cannot explain the appearance of an extra feature above the bunch of Ni related bands as observed in ARPES.}


\MM{Finally, we visualize the expected 3d level occupation for NiPS$_3$ and MnPS$_3$  by using the octahedral coordinates along the bond directions to assign the 3d level orbitals (Fig.~\ref{Fig_4}g-l).} For Mn$^{2+}$ 3d$^5$, a complete spin polarization of the 3d levels is obtained (Fig.~\ref{Fig_4}k) as expected from Hund's rule (Fig.~\ref{Fig_4}l). In addition, the occupied e$_{\rm g}$ levels are split into two parts due to the deviations from a perfect octahedral symmetry \cite{Figgis2000,Trimarchi2018}. 
On the other hand, 
Ni$^{2+}$ 3d$^8$ shows completely occupied t$_{2g}$ levels, while only the e$_g$ levels are close to fully spin-polarized (Fig.~\ref{Fig_4}h). 
Hence, the occupation of bands is in line with the expected 3d level occupations and energy splittings. Moreover, for both materials, the bands that are changing most strongly across $T_{\rm N}$ exhibit significant 3d contributions. This makes us confident that the band assignment by DFT+U is correct and the feature at $E-E_{\rm F}^{\rm Mn}\approx -1$\,eV is beyond the related approximations. It is very likely that 3d level multiplets are involved, maybe including ligand contributions \cite{Kang2020}, as will be discussed in a separate publication.

\section{Conclusions}

We probed the band structure of NiPS$_3$ above and below the Néel temperature using $\mu$-ARPES on flakes exfoliated onto a Au surface. The ARPES data are compared with DFT+U calculations to identify the orbital character of bands. We identify an  upwards band shift  below $T_{\rm N}$ by roughly 150\,meV for a band close to $\overline{\Gamma}$. This band consists of mixed Ni 3d and S 3p orbitals that likely feature an antiferromagnetic Ni 3t$_{\rm 2g}$-S 3p-Ni 3t$_{\rm 2g}$ superexchange path.
%
Beyond the good agreement between DFT+U and ARPES that we also found for FePS$_3$ \cite{Pestka2024} and MnPS$_3$ \cite{Strasdas2023}, we identified a shoulder in the ARPES data above the valence band maximum that could not be attributed to a band of the DFT+U data. This is in striking contrast to FePS$_3$ and MnPS$_3$. We assume that the additional structure is reminiscent of the known strong correlations in NiPS$_3$.

\MMM{{\it Note added in proof}: A very recent publication also shows $T$-dependent ARPES data of NiPS$_3$ revealing similar band structure features in reasonable correspondance with a similarly selected $U=2.0$\,eV \cite{Cao2025}.}

\section*{Materials and Methods}

\textit{NiPS$_3$ crystal growth}: Bulk single crystals of NiPS$_3$ were synthesized via the vapor-transport method using stoichiometric amounts of Ni, red P, and S powders, with an additional 5$\%$ excess of sulfur to act as a transport agent. The mixture was ground to a fine powder and then sealed under high vacuum ($\sim 4\cdot10^{-5}$ mbar) in a quartz ampoule. This ampoule was then placed in a two-zone furnace, set at 740 $^{\circ}$C in the precursor zone and 690 $^{\circ}$C in the product zone, for one week. \\

\textit{Sample preparation}: We employ an optimized exfoliation method (scotch tape: Nitto ELP BT-150E-KL) to prepare samples with few-layered flakes of NiPS$_3$ in a cleanroom environment. Commercially available Si/SiO$_2$ (oxide thickness 90\,nm) is used as substrate. It has been metal-coated at 370\,K with Au (5\,nm)/Ti (1\,nm) in an ultrahigh vacuum e-beam evaporator. This improves the flake adhesion and renders the substrate conducting, which is necessary to prevent charging during ARPES \cite{Strasdas2023,Koitzsch2023,Velicky2018}. A mild O$_2$ plasma ashing (power = 50\,W, flux = 100\,sccm) is done at room temperature for 25\,s just before exfoliating NiPS$_3$ flakes 
to efficiently remove organic molecules from the gold surface by oxidation. 
NiPS$_3$ is exfoliated at 60$^{\circ}$C within few minutes after the plasma cleaning. This leads to flakes with diverse thickness, where the desired thickness and flake size are identified using commercial optical and atomic force microscopes \cite{Strasdas2023}. \\

\textit{ARPES and XPS experiments}: $\mu$-ARPES and XPS were performed at the NanoESCA beamline of the Elettra synchrotron radiation facility in Italy. An ultrahigh vacuum FOCUS NanoESCA photoemission electron microscope (PEEM) was used for the $k$-space mapping mode operation working at a background pressure of $5\cdot 10^{-11}$\,mbar \cite{Schneider2012}. To ensure clean surfaces, the samples are firstly annealed at 200$^{\circ}$C for a couple of hours prior to ARPES. The synchrotron allows to focus a wide range of photon energies on the sample with a beam spot of 5-10\,$\mu$m in diameter and total energy resolution (beamline and analyzer) of 50\,meV. For our experiments, the photon beam impinged with an incident angle of $65^\circ$ with respect to the surface normal and with a $5^\circ$ azimuth relative to the $\overline{\rm M}\overline{\Gamma}\overline{\rm M}$ direction of the crystal. The polarization was within the plane of incidence (p-polarized). The Fermi level has been determined by ARPES on the Au film of the substrate. The sample temperature is adjusted by changing the liquid He flow rate and measured by a Lake Shore DT-670E-BR Si diode placed at the sample holder. \MM{So far, only two temperatures are employed for technical reasons such that ARPES data at temperatures closer to $T_{\rm N}$ have to be postponed to future experiments.}  \\



 

\textit{Computational Details}: 
The calculations were performed in the framework of density functional theory (DFT) using the generalized gradient approximation within the PBE  flavor \cite{PhysRevLett.77.3865}, as implemented in VASP software \cite{Kresse1996}. The ion–electron interactions were described by the projector augmented wave (PAW) method  \cite{Holzwarth2001}. Plane-wave basis cutoff and $\Gamma$ centered Monkhorst-Pack \cite{Monkhorst1976} k-point grid were set to 550 eV and $8\times14\times8$, respectively. A Gaussian smearing of 0.05 eV was employed for the Brillouin zone (BZ) integration. The interlayer vdW forces were treated within the Grimme scheme using D3 correction \cite{doi:10.1063/1.3382344}. All of the results were obtained  using PBE+U method based on Dudarev's approach \cite{PhysRevB.57.1505}, with the effective on-site Hubbard U parameter ($U_{\rm eff} = U-J$, where $J$ is fixed to $J=1$\,eV) for the 3d orbitals. 
The position of the atoms and unit cell were fully optimized within the PBE+U approach. To simulate the disordered paramagnetic state, we used a $4\times 2\times 3$ supercell with spin-orbit interaction and randomly set the directions of the magnetic spin moments (in 3D) so that the net moment was zero. Here, we did not optimize the geometric structure, in order to capture only changes in electronic dispersion. 
Moreover, we used the widely employed band unfolding method to obtain an effective electronic structure \cite{Popescu2012}. \MMM{Details on additional calculations given in the Supporting Information are provided in Suppl. Section S13.}

\section*{Associated Content}
\noindent \textbf{Data Availability Statement}\\
All data that support the findings of this study are available in zenodo at DOI 10.5281/zenodo.16162323.\\  

\noindent \textbf{Notes}\\
A preprint of this manuscript has been submitted to arXiv~\cite{Pestka2025_arxiv}.\\
The authors declare no competing financial interest.


\section*{Supporting Information}
\MM{A possible application of insulating magnetic 2D materials}, details on sample preparation, experimental and computational methods, adaption of $U_{\rm eff}$ and $k_z$ including $h\nu$ dependence of ARPES data, individual orbital projections of bands, additional charge density plots, ARPES and DFT band structure \MM{along distinct $\overline{K}\overline{\Gamma}\overline{K}$ directions}, DFT+U band structure of paramagnetic phase, analysis of ARPES structure at $E-E_{\rm F}=-1.3$\,eV, \MM{detailed description of the use of ARPES curvature and additional DFT calculations employing HSE and using GW methods.}\\

\section*{Acknowledgement}

We acknowledge financial support from the German Research Foundation (DFG) via the project Mo 858/19-1 and Germany’s Excellence Strategy Cluster of Excellence: Matter and Light for Quantum Computing(ML4Q) EXC 2004/1 (project Nr. 390534769) from the German Ministry of Education and Research (Project 05K2022 -ioARPES).
B.B. acknowledges the support of Humboldt Research Fellowship funded by the Alexander von Humboldt Foundation. 
M.B. acknowledges financial support from the National Science Centre (NCN), Poland [2024/53/B/ST3/04258]. We gratefully acknowledge Polish high-performance computing infrastructure PLGrid (HPC Center: ACK Cyfronet AGH) for providing computer facilities and support within computational grant no. PLG/2024/017490. K.W. acknowledges support by the  National Science Centre in Poland under the project 2024/55/B/ST3/03144.
A.K.B. and E.L. were supported by the European Commission via the Marie-Sklodowska Curie action Phonsi (H2020-MSCA-ITN-642656).

\bibliography{references}
\end{document}


\title{Supplementary Information: Probing the band structure of the strongly correlated antiferromagnet NiPS$_3$ across its phase transition}

\author{Benjamin Pestka$^\dag$}\affiliation{\AC}
\author{Biplab Bhattacharyya$^\dag$}\affiliation{\AC}
\author{Miłosz Rybak$^\dag$}\affiliation{\PW}
\author{Jeff Strasdas}\affiliation{\AC}
\author{Adam K. Budniak}\affiliation{\Haifa}
\author{Adi Harchol}\affiliation{\Haifa}
\author{Marcus Liebmann}\affiliation{\AC}
\author{Niklas Leuth}\affiliation{\AC}
\author{Honey Boban}\affiliation{\FZJ}
\author{Vitaliy Feyer}\affiliation{\FZJ}
\author{Iulia Cojocariu}\affiliation{\FZJ}\affiliation{\Trieste}\affiliation{\Elettra}
\author{Daniel Baranowski}\affiliation{\FZJ}\affiliation{\PNNL}
\author{Simone Mearini}\affiliation{\FZJ}
\author{Lutz Waldecker}\affiliation{\ac}
\author{Bernd Beschoten}\affiliation{\ac}
\author{Christoph Stampfer}\affiliation{\ac}
\author{Yaron Amouyal}\affiliation{\HaifaMat}
\author{Lukasz Plucinski}\affiliation{\FZJ}
\author{Efrat Lifshitz}\affiliation{\Haifa} 
\author{Krzysztof Wohlfeld}\affiliation{\IFT}
\author{Magdalena Birowska}\affiliation{\IFT}
\author{Markus Morgenstern$^*$}\affiliation{\AC}
\date{\today} 

\maketitle
\noindent {{$^\dag$}These authors contributed equally to this work.
{$^*$}Corresponding author: M.~Morgenstern, Email: \href{mmorgens@physik.rwth-aachen.de}{mmorgens@physik.rwth-aachen.de}} 

\tableofcontents
\FloatBarrier
\section{Possible electron-magnon transducer}
\BB{The strong exciton–magnon coupling in van der Waals magnets as NiPS$_3$ offers novel opportunities for optomagnetic applications, e.\,g.  towards transducers between magnons and photons in the infrared range. It has been shown that magnons
can be tracked coherently in 2D materials, e.\,g. in CrSBr, by time-resolved shifts of their exciton energy by $\Delta E\sim 20$\,meV down to ps time scales~\cite{Bae2022}. More recently, pump–probe spectroscopy on NiPS$_3$ revealed ultrafast reflectivity changes that are related to spin dynamics via the critical slowing down around $T_{\rm N}$~\cite{Sahu2025}.  
On the other hand, magnons as information carriers have been intensely pursued, meanwhile leading to rather complex devices as, e.\,g., spin-wave multiplexers~\cite{Vogt2014, Au2012, Zhang2014}. Thus, magnon–exciton coupling, in which Coulomb-bound electron–hole pairs interact with the collective spin excitations, could enable transducers between spin information and optical channels or vice versa. Therefor, the magnon properties of the 2D materials can be imprinted coherently in the optical intensity as already demonstrated \cite{Diederich2022, Bae2022}, or into the light polarization since the exciton’s dipole moment of, e.\,g. NiPS$_3$ is locked to the local spin orientation \cite{Wang2021}. 
Moreover, the exciton-magnon interaction in 2D materials is tunable via strain or interface design, shifting magnon frequencies or adjusting coupling strengths \cite{Diederich2022}. This is favorable for tunable or selective transducers. Finally, NiPS$_3$ has a relatively large fundamental magnon gap of 5.3\,meV with narrow linewidth (0.1 meV) as probed by THz absorption ~\cite{Belvin2021}. This points towards low noise magnonic operation. The transfer of the magnons, e\,g. to Y$_3$Fe$_5$O$_12$ (YIG) still has to be explored, but is possible in principle \cite{Wang2014}. Such an approach towards efficient magnon-photon transducers in the infrared would combine the low-dissipation propagation of magnons with the coherent long-range communication by photons via fibers, eventually enabling more scalable, magnonic technologies. Of course, the relatively efficient spin-optical interface might also be used for other interconnects between spin-based devices.} 

\section{Adjusting the Brillouin zone to the ARPES data}

\begin{figure}[thb]
\centering
\includegraphics[width=0.5\textwidth]{Fig_S1.pdf}
\vspace{0 cm}
\caption{{\bf Brillouin zone (BZ)} (a) Two-dimensional projections of the BZ to the surface for the atomic arrangement of NiPS$_3$ without magnetism (grey hexagon) and for the zigzag-type AFM arrangement (blue rectangle), high symmetry points are marked. 
(b) (\textit{k}$_\textit{x}$, \textit{k}$_\textit{y}$) plot of ARPES curvature intensity, \textit{E}-\textit{E}$_\text{F}$ = -7.65\,eV, \textit{h}$\nu$ = 60\,eV, $T = 45$\,K, the hexagonal surface projection of the atomic BZs is overlaid (dashed white lines), arrows: $\overline{\Gamma}\overline{\rm M}$, $\overline{\Gamma}\overline{\rm K}$ direction. 
}
\label{Fig. S1}
\end{figure}

\begin{figure*}[htb]
\centering
\includegraphics[width=\textwidth]{Fig_S2.pdf}
\vspace{-0.5 cm}
\caption{{\bf Comparing folded and unfolded DFT+U data.} (a) DFT+U band structure displayed in the magnetic Brillouin zone (blue rectangle) with all bands, $U_{\rm eff}=1.6$\,eV, \textit{k}$_\textit{z}= 0.1/$\AA, symbol diameter is proportional to the percentage of the orbital contribution to this state, color code as in Fig.~3, main text. (b) Same as a after unfolding the bands to the hexagonal atomic BZ (gray hexagon) \cite{Popescu2012}. 
(c), (d) Same as a and b, but depicting only the selected orbitals as marked on top according to simplified selection rules \cite{Moser2017}.
}
\label{Fig. S3}
\end{figure*}

\begin{figure*}[thb]
\centering
\includegraphics[width=\textwidth]{Fig_S3.pdf}
\vspace{-0.5 cm}
\caption{{\bf Adapting the $U_{\rm eff}$ parameter.} (a) ARPES curvature plot, \textit{h}$\nu$ = 60\,eV, $T = 45$\,K, $\overline{\rm M}\overline{\Gamma}\overline{\rm M}$ direction, same color scale as in Fig.~3, main text. (b)-(g) DFT+U band structure at various $U_{\rm eff}$ as marked for selected orbitals with s, p$_z$ and d$_{z^2}$ character, symbol diameter is proportional to the percentage of the orbital contribution to this state, color code as in Fig.~3, main text, \textit{k}$_\textit{z}= 0.1/$\AA. Labels i-ix and dashed blue arrow are used to compare theory with experiment as discussed in the text. The red box highlights the selected $U_{\rm eff} = 1.6$\,eV.
}
\label{Fig. S2}
\end{figure*}

Figure~\ref{Fig. S1}a displays the Brillouin zone projection to the surface (Fig.~\ref{Fig. S5}a) 
for the atomic arrangement without magnetism (grey hexagon) and with the antiferromagnetic (AFM) zig-zag structure (blue rectangle). The $\overline{\rm K}$ points are backfolded to the interior of the magnetic BZ, while the $\overline{\rm M}$ points are either not affected (four times) or backfolded to $\overline{\Gamma}$ (two times). In the main text, we consistently used the atomic, hexagonal Brillouin zone for labeling and we unfolded to this Brillouin zone for the DFT+U data \cite{Popescu2012}.   
To adjust the BZ orientation to the angular dependence of the photoelectron intensity, we employ ($k_x,k_y$) plots  at energies where bands exhibit maxima at the BZ boundary. For feature ix of the band structure (mixed P/S band, Fig.~\ref{Fig. S2}a,d), this reveals a hexagonal symmetry  (Fig.~\ref{Fig. S1}b) that can be adapted to the surface projection of the atomic BZ (dashed hexagons). This symmetry is visible above and below $T_{\rm N}$ showing that the ARPES data of this band are not affected by the magnetically induced backfolding. We checked that this is true for all energies where the hexagonal symmetry is clearly apparent, including bands that involve Ni 3d orbitals.
Hence, the AFM superstructure does not change the periodicity of the wave functions significantly as also evidenced by the unfolding procedure (Fig.~\ref{Fig. S3}). The overlay of the hexagon edges of the projected BZ to the maxima in the ($k_x,k_y$) plot reveals straightforwardly the orientation of the BZ and the relation between photoemission angle and the value of $k_\parallel$     
by relating to values from the literature ($\overline{\Gamma}\overline{\rm M}: 0.62/$\AA, $\overline{\Gamma}\overline{\rm K}: 0.72/$\AA\, \cite{Ouvrard1985,Bernasconi1988}).

Figure~\ref{Fig. S3} displays the comparison between the calculated initial states in the magnetic, rectangular Brillouin zone and the unfolded data (hexagonal Brillouin zone), once for all bands (Fig.~\ref{Fig. S3}a-b) and once for only the bands with more than 10\,\%  contribution of s, p$_z$ and d$_{z^2}$ orbitals (Fig.~\ref{Fig. S3}c-d). We choose a $\overline{\rm M\Gamma M}$ direction that is unchanged by the AFM order (green double arrow in the Brillouin zone at the upper left) to ease the comparison. Multiple bands are simply backfolded by the AFM periodicity and disappear by unfolding. Importantly, this reduces the relevant bands for comparison to the ARPES data significantly besides the reduction to orbital contributions given by the matrix element projection to final state plane waves as for our electric field geometry of the incident photon beam \cite{Moser2017,Strasdas2023}. 



\section{Adapting $U_{\rm eff}$ of DFT+U calculations to ARPES data}
\label{sec:U_select}

In order to assign orbital characters to the bands visible in  ARPES, one has to select the best matching on-site energy $U_{\rm eff}$ of the DFT+U calculations as well as the best matching wave number $k_z$ perpendicular to the surface. For that purpose, we employ several pronounced features of the ARPES data and recursively compare them with the DFT+U band structure for different $U_{\rm eff}$ and $k_z$.  We eventually selected $U_{\rm eff}=1.6$\,eV and $k_z=0.1$/\AA\, as described in detail in the next sections. This selection does not lead to a one-to-one correspondence between ARPES and DFT+U data, but it gives a solid understanding of the orbital development of the bands with these parameters and, hence, of its robustness.    
In Fig.~3, main text, the most important features in ARPES are already labeled as i-ix. For comparison of these features with DFT+U band structures, we only use the bands with a strong s, p$_z$ or d$_{z^2}$ character according to the simplified selection rules in our photon beam geometry as discussed in the main text \cite{Moser2017} and the unfolded band structure.

Figure~\ref{Fig. S2} shows the comparison of ARPES curvature data, recorded at \textit{h}$\nu$ = 60\,eV,  with DFT+U data at various $U_{\rm eff}$ and optimized $k_z=0.1$/\AA.  
The lowest energy band  (feature ix) that is nearly parabolic at $\overline{\Gamma}$ and consists of S 3p and P 3p contributions (Fig.~\ref{Fig. S6}) barely shifts with energy due to its missing Fe 3d orbitals. We use it to rigidly shift the energy scale of the DFT+U data 
to the ARPES data given with respect to $E_{\rm F}$. The band from DFT+U is chosen slightly below the band in ARPES for reasons described latter. The rigid shift implies a slight n-doping of the material regarding the known energy gap of 1.8\,eV in NiPS$_3$ \cite{Kim2018}. This is visible as faint bands at the top of the calculated band structures belonging to the conduction band. 

\begin{figure*}[thb]
\centering
\includegraphics[width=\textwidth]{Fig_S4.pdf}
\vspace{-0.5 cm}
\caption{{\bf Adapting \textit{k}$_\textit{z}$ parameter.} (a) ARPES curvature plot, \textit{h}$\nu = 60$\,eV, $T = 45$\,K, $\overline{\rm M}\overline{\Gamma}\overline{\rm M}$ direction, same color scale as in Fig.~3, main text. (b)--(g) DFT+U band structures at various \textit{k}$_\textit{z}$ as marked, $U_{\rm eff} = 1.6$\,eV, only bands with significant s, p$_z$ and d$_{z^2}$ character are displayed \cite{Moser2017}, symbol diameter is proportional to the percentage of the orbital contribution to this state, color code as in Fig.~3, main text. Labels i-ix and dashed blue arrow are used for comparison to ARPES as discussed in the text. The red box highlights the selected \textit{k}$_\textit{z}$ = 0.1/\AA.
}\label{Fig. S4}
\end{figure*}

For $U_{\rm eff}$ selection, we use bands that are changing with $U_{\rm eff}$. 
The topmost flat green bands (feature i) of dominating S 3p character with contributions from Ni 3d$_{xz,yz}$ (Fig.~\ref{Fig. S6}) move downwards with increasing $U_{\rm eff}$ matching the experiment at $U_{\rm eff}=1.6-2.0$\,eV.
Feature ii is a relatively flat band in ARPES identified with the flat DFT+U band of dominating Fe 3d$_{z^2}$ character. It also moves downwards with energy matching best at $U_{\rm eff}=2.0$\,eV, where also the gap between features i and ii matches the experiment. The small gap at $\overline{\Gamma}$ between features iv and vi appears reminiscent to a gap opening at a former crossing point in the experiment. This gap is most pronounced at $U_{\rm eff}=1.0-1.2$\,eV in DFT +U being more faint at $U_{\rm eff}=1.6$\,eV and matches best to the experiment at $U_{\rm eff}=1.2$\,eV.  The bands surrounding the crossing point at $\overline{\Gamma}$ get, moreover, significantly more flat than in the experiment for $U_{\rm eff}>1.6$\,eV. Recall that this is the area of band change across $T_{\rm N}$ (Fig.~2, main text). The steep band between features vi and vii is not found continuous in energy for any $U_{\rm eff}$, but exhibits a reasonably similar slope as in the ARPES data around vii for $U_{\rm eff}=1.0$\,eV with continuously decreasing slope for higher $U_{\rm eff}$. Feature viii, finally, is too weak in experiment and calculations to be used for $U_{\rm eff}$ selection.

Summarizing, $U_{\rm eff}=1.6$\,eV is the best compromise between the slightly larger $U_{\rm eff}$ required for the high energy bands and the slightly lower $U_{\rm eff}$ necessary to describe the lower energy bands.

\section{Adapting $k_z$ of DFT+U calculations to ARPES}
\label{sec:kz_select}

\begin{figure*}[!thbp]
\centering
\includegraphics[width=1\textwidth]{Fig_S5.pdf}
\vspace{-0.3 cm}
\caption{{\bf Photon energy ($k_z$) 
variation.}
(a) Sketch of the atomic bulk BZ of NiPS$_3$ (black) with 2D surface projection on top (orange) as well as corresponding cross section (orange). High symmetry points and $k_\parallel$ directions (red lines) are marked. 
The upper/lower bulk BZ boundary is at \textit{k}$_\textit{z} = \pm 0.48$/\AA. 
(b) (\textit{k}$_\textit{x}$, \textit{k}$_\textit{y}$) plot of ARPES curvature, \textit{E}-\textit{E}$_\text{F}$ = -7.65\,eV, \textit{h}$\nu$ = 60\,eV, $T = 45$\,K, hexagonal surface projections of the atomic BZ boundaries are overlaid (dashed white lines)  (same as Fig.~\ref{Fig. S1}b). 
(c)-(g) ARPES curvature  at various \textit{h}$\nu$ as marked on top, $\overline{\rm M}\overline{\Gamma}\overline{\rm M}$ direction, $T = 45$\,K, \textit{k}$_\textit{z}$ values maked on top are deduced using the free electron final state model (eq.~(\ref{eq:S1})) with inner potential V$_0$ = 12.1\,eV at $E-E_{\rm F}\approx -3.7$\,eV. 
The $k_z$ variation in each plot is $\Delta k_z \approx 0.25/$\AA\, with the largest value at the top. (h)-(l) Band structures from DFT+U at the $k_z$ marked below, close to the $k_z$ value of the ARPES data on top, $U_{\rm eff}=1.6$\,eV, only s, p$_z$, d$_{z^2}$ orbitals, colorcode as in Fig.~3, main text. 
The blue, dashed rectangles mark a similar \textit{k}$_\textit{z}$ dispersion of feature ix (P/S band). (c)-(l) share the same energy scale as displayed in c.
}
\label{Fig. S5}
\end{figure*}

We followed a similarly detailed comparison as described in Supplementary Section~\ref{sec:U_select} for matching \textit{k}$_\textit{z}$. Figure~\ref{Fig. S4} depicts the ARPES data and the DFT+U data at $U_{\rm eff}=1.6$\,eV for various $k_z$ highlighting the eventually selected $k_z=0.1/$\AA. One has to keep in mind that $k_z$ is changing for constant $h\nu$ with $E-E_{\rm F}$ by roughly 0.25/\AA\, across the energy range displayed in Fig.~\ref{Fig. S4}a (see below) with the largest $k_z$ at the top.  

We start with features i and ix, which move downwards and upwards, respectively, with increasing $k_z$. The best match of their energy distance to the ARPES data is at $k_z\approx0.3$/\AA. Next, the gap between features i and ii gets smaller with increasing $k_z$ and matches best to the ARPES data at $k_z=0.0-0.2$/\AA. Moreover, the wing-like features consisting of iii, iv and v change in shape with $k_z$ exhibiting a reasonable match at $k_z=0.1$/\AA. This is also true for the steepness of the bands around feature vi that are less steep at higher and lower $k_z$ than $k_z=0.1$/\AA. Finally, the suppressed curvature between features ii and iv might be related to the only weakly appearing bands at $k_z \ge 0.1$/\AA\, in that energy range, where bands are stronger at $k_z=0.0$/\AA. Hence, with a certain 
ambiguity, we choose $k_z = 0.1$/\AA\, as the best compromise.

Using the $k_z$ selection for $h\nu=60$\,eV, we  can relate different photon energies to different $k_z$. We use the simplified free electron model for final states in the crystal with origin of the corresponding free electron parabola at $V_0$ \cite{Hüfner2013}. 
Hence, the final state energy $E_{\rm final}$ at wavevector $\textbf{\textit k}$ inside the crystal reads:  
\begin{equation}
\label{eq:S1}
E_{\rm final}(\textbf{\textit{k}}) = \frac{\hbar^2\textbf{\textit k}^2}{2m}-V_0. 
\end{equation}
%
The kinetic electron energy in vacuum reads: 
\begin{equation}
\label{eq:S2}
E_\text{kin}^\text{vac} = E_\text{kin}^\text{A} - \phi + \phi^\text{A}
\end{equation}
with the work function of the NiPS$_3$ flake $\phi=5.375$\,eV, the work function of the analyzer $\phi^\text{A}=4.6$\,eV, and the kinetic energy $E_{\rm kin}^{\rm A}$ of the photoelectrons at the analyzer. $\phi^\text{A}$ was determined using $E^{\rm A}_{\rm kin}$ of photoelectrons emitted from the Fermi level $E_{\rm F}$ measured on the gold substrate next to a NiPS$_3$ flake. $\phi$  is determined on a NiPS$_3$ flake from the onset energy of the secondary photoelectrons, i.\,e. from the minimum value of the free electron parabola of onset energies of secondary electron emission as a function of emission angle that is determined by a parabolic fit \cite{Strasdas2023}.

\begin{figure*}[thb]
\centering
\includegraphics[width=\textwidth]{Fig_S6.pdf}
\vspace{-0.5 cm}
\caption{{\bf Orbital projections of bands.} (a) ARPES curvature, \textit{h}$\nu = 60 $\,eV, $T = 45$\,K, $\overline{\rm M}\overline{\Gamma}\overline{\rm M}$ direction, color scale as in Fig.~3, main text. (b)-(g) DFT+U band structure with selected orbitals as marked on top, $U_{\rm eff} = 1.6 $\,eV, \textit{k}$_\textit{z} = 0.1$/\AA, symbol diameter is proportional to the percentage of the orbital contribution to this state, color code as in Fig.~3, main text.  Labels i-ix mark  features discussed in the text.
}
\label{Fig. S6}
\end{figure*}

To estimate $V_0$, we use the photoelectron intensity of the most relevant feature iv for $h\nu = 60$\,eV and emission angle $\theta = 0^\circ$. It appears at $E_\text{kin}^\text{A}=51.7$\,eV and is related to $k_\parallel=0.0/$\AA\ and $k_z=0.1/$\AA\ in the crystal as deduced from the comparison with the DFT+U data (Fig.~\ref{Fig. S2}, \ref{Fig. S4}). Adequate rearrangement of eq.~(\ref{eq:S1}) and eq.~(\ref{eq:S2}) eventually results in \cite{Sobota2021}:
\begin{equation}
\label{eq:S3}
V_0 = \frac{\hbar^2}{2m} (n G_\perp+k_z)^2 - E_\text{kin}^\text{A} + \phi - \phi^\text{A},
\end{equation}
where $G_\perp = 0.9917/$\AA\ is the reciprocal lattice vector of NiPS$_3$ normal to the surface \cite{Brec1986} and $n$ is an integer.  The free choice of $n$ gives an uncertainty in deducing $V_0$.
At $n=4$, eq.~(\ref{eq:S3}) results in the lowest positive value of $V_0 \approx 12.1\,$eV, while, for $n=5$, one obtains an unrealistically large $V_0 \approx 46.6$\,eV \cite{Zhang2022}. Thus, $n=4$ is the most reasonable. Inserting it into eq.~(\ref{eq:S3}) and solving for $k_z$ reveals the values for other photon energies at $\theta=0^\circ$ and $E-E_{\rm F}=-3.7$\,eV \cite{Sobota2021} as marked in Fig.~\ref{Fig. S5}c-g. The $k_z$ value obviously varies with $E-E_{\rm F}$, respectively $E_\text{kin}^\text{A}$ for constant $h\nu$. Within the displayed $E-E_{\rm F}$ range of Fig.~\ref{Fig. S5}c-g, $k_z$ changes by roughly 0.25/\AA\  (25\,\% of the BZ) with the largest $k_z$ at the top of the images.

\begin{figure*}[!thbp]
\centering
\includegraphics[width=0.85\textwidth]{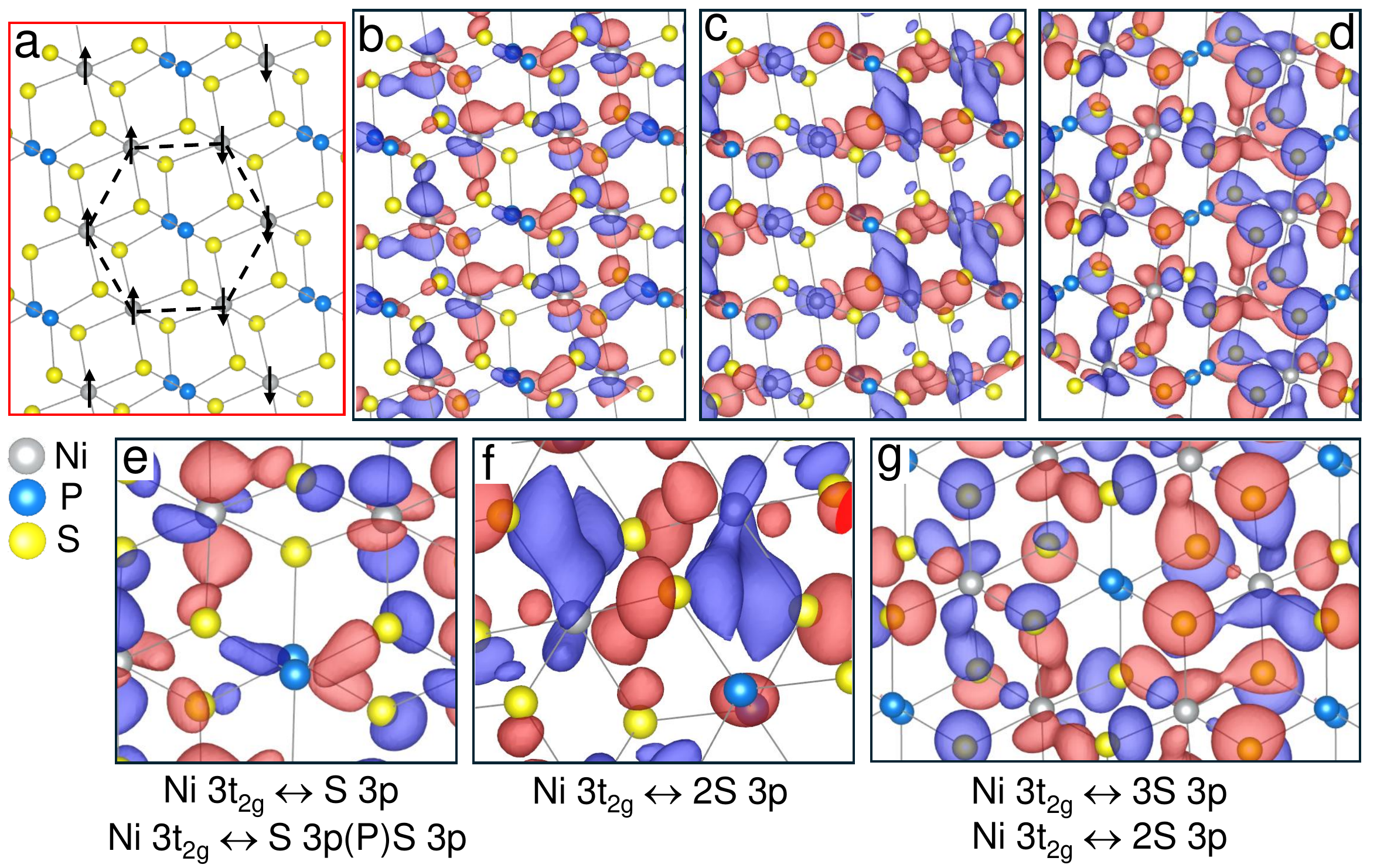}
\vspace{-0.3 cm}
\caption{\textbf{Charge density of states from the magnetically modified energy range.} (a) Crystal structure of a single NiPS$_3$ layer viewed along the c axis with a marked honeycomb of six Ni atoms (dashed line) and spin directions in the zigzag AFM configuration (black arrows), same as Fig.~4a, main text. (b)--(d) Contour planes of the charge density $|\psi(\boldsymbol{x})|^2$ of three distinct states at $E-E_{\rm F}\approx -3.9$\,eV and $\boldsymbol{k}=\boldsymbol{0}$/\AA\, (marked by an arrow in Fig.~3c, main text). The area and angle of view are the same as in a. Blue and red contour planes mark opposite signs of the wave function. (e)--(g) Zoom into b--d, respectively, at an optimized angle of view showing the Ni bonds and orbitals more clearly. The identified bonds are indicated below (graphics made by VESTA \cite{VESTA}). 
}
\label{FigS7a}
\end{figure*}

\begin{figure}[!b]
\centering
\includegraphics[width=0.42\textwidth]{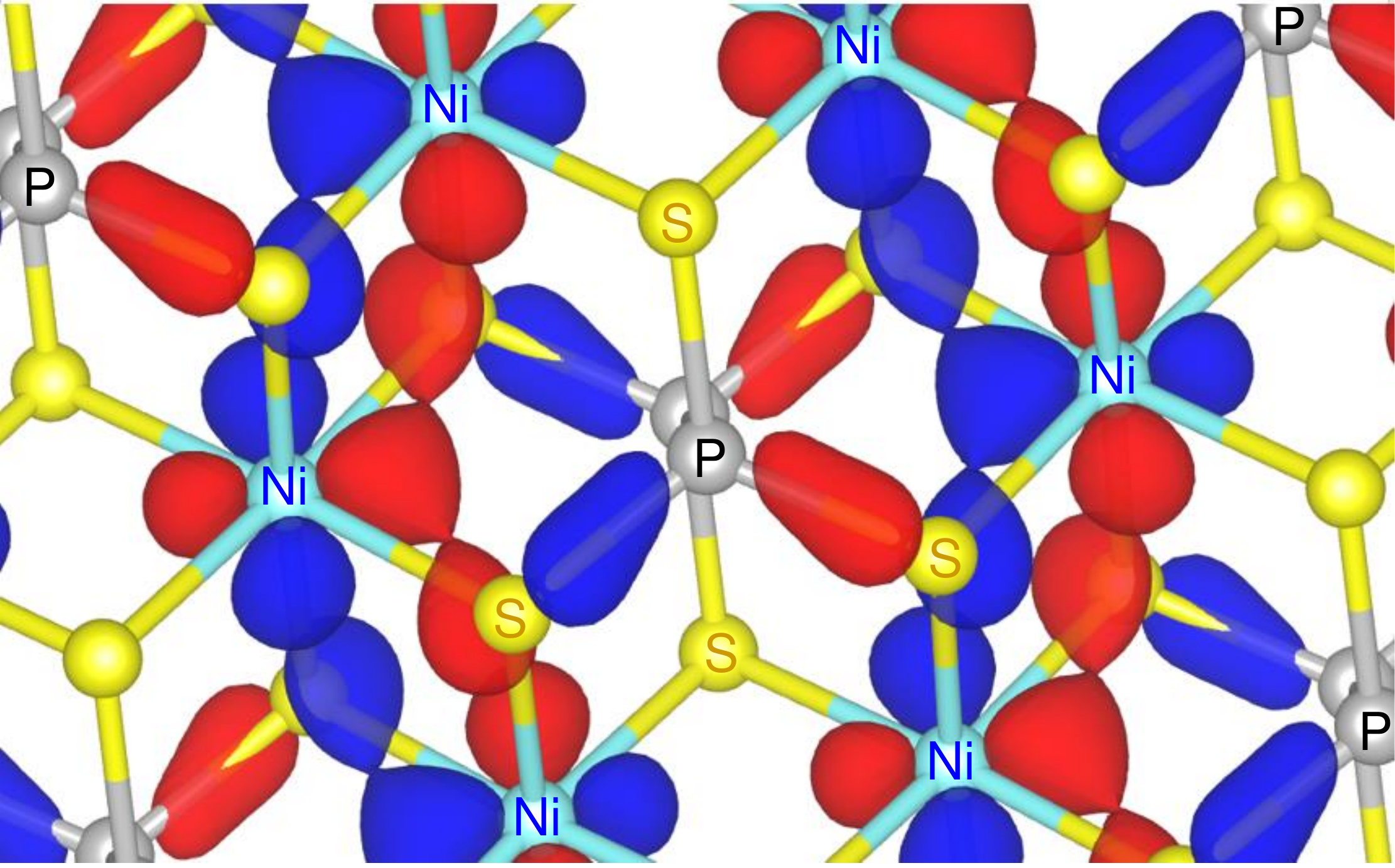}
\vspace{-0.3 cm}
\caption{\textbf{Charge density of shifting band in Néel configuration.}  Contour planes of the charge density $|\psi(\boldsymbol{x})|^2$ of a state at $E-E_{\rm F}=-3.9$\,eV and $\boldsymbol{k}=\boldsymbol{0}$/\AA\, in the Néel configuration with labeled atoms (zoom view, graphics made by VESTA \cite{VESTA}). 
}
\label{FigS7b}
\end{figure}

%

\begin{figure*}[bth]
\centering
\includegraphics[width=0.9\textwidth]{Fig_S9.pdf}
\vspace{-0.5 cm}
\caption{{\bf Band structure in $\overline{\rm K}\overline{\Gamma}\overline{\rm K}$ and $\overline{\rm M}\overline{\Gamma}\overline{\rm M}$ direction.} (a), (d) ARPES curvature along $\overline{\rm K}\overline{\Gamma}\overline{\rm K}$ and $\overline{\rm M}\overline{\Gamma}\overline{\rm M}$ directions as marked on top, \textit{h}$\nu = 60$\,eV, $T = 45$\,K. (b), (c) DFT+U band structure of only the s, p$_z$ and d$_{z^2}$ contributions, $U_{\rm eff} = 1.6 $\,eV, \textit{k}$_\textit{z} = 0.1$/\AA, symbol diameter is proportional to the percentage of the orbital contribution,  color code as in Fig.~3, main text.  
}
\label{Fig. S7}
\end{figure*}

As a rough crosscheck of the $k_z$ selection, we display the $k_z$ dependence of the DFT+U data
directly below the $h\nu$ dependent ARPES data. The $k_z$ values are not perfectly adapted due to the finite $\boldsymbol{k}$ point density in the calculation, but since $k_z$ varies with $E-E_{\rm F}$ at constant $h\nu$ anyhow, this does not matter. Favorably, several trends with $k_z$ are found similarly in calculation and experiment. Firstly, feature ix (blue dashed rectangle) moves upwards and gets less dispersive for $k_z\ge 0.2$/\AA, both in ARPES and DFT+U data. Secondly, feature ii, corresponding to Ni 3d$_{z^2}$ bands, moves upwards  and gets  more intense in dispersion from left to right, both for experiment and calculation.  For negative $k_z$, the dispersion
of the Ni 3d$_{z^2}$ band ii is even largely identical between experiment and calculation. Thirdly, at negative $k_z$, there is a flat Ni 3d band below feature vi (Fig.~\ref{Fig. S5}h,i) that appears surprisingly similar in the experimental data (Fig.~\ref{Fig. S5}c,d). Note that the in-plane AFM structure of NiPS$_3$ breaks time-reversal symmetry for the $k_z$ direction and, hence, enables different $E(k_\parallel)$ at positive and negative $k_z$ for the same $|k_z|$, as found in ARPES and DFT+U.

 There are also some discrepancies such as the upward movement of feature i with increasing $k_z$  in the ARPES data, but not in the DFT+U data for an unknown reason. Nevertheless, $k_z=0.1$/\AA\, appears to be a favorable choice for $h\nu = 60$\,eV at $E-E_{\rm F}=-3.7$\,eV, i.\,e. the band energy region with changes across $T_{\rm N}$.  

Obviously, there is considerable $k_z$ dispersion of the electronic band structure, both in the ARPES and the DFT+U data, revealing a sizable electronic interaction between the layers. 
Changes with $h\nu$ in the ARPES data might be partly due to changing matrix elements, but this is not the case for the $k_z$ dependence of the DFT+U band structure.

\begin{figure*}[bth]
\centering
\includegraphics[width=1.0\textwidth]{Fig_Sx1.pdf}
\vspace{-0.4 cm}
\BB{\caption{{\bf Comparison of DFT+U band structure along non-equivalent $\overline{\rm K}\overline{\Gamma}\overline{\rm K}$ directions.} 
(a) Scheme of the hexagonal 2D Brillouin zone with color coded $\overline{\rm K}\overline{\Gamma}\overline{\rm K}$ directions. (b)-(d) DFT+U band structure along the three directions highlighted in a as marked by the $\boldsymbol{k}_\parallel$-axis symbol colors, only s, p$_z$ and d$_{z^2}$ contributions, $U_{\rm eff} = 1.6 $\,eV, \textit{k}$_\textit{z} = 0.0$/\AA, symbol diameter is proportional to the percentage of the orbital contribution,  color code as in Fig.~3, main text. Minor differences are highlighted by arrows. 
(e)-(g) Same as b-d, but for \textit{k}$_\textit{z} = 0.1$/\AA.
The asymmetry between the positive and the negative $\overline{\Gamma}\overline{\rm K}$ directions are due to the missing mirror symmetry in the Brillouin zone planes perpendicular to $k_z$ (Fig.~\ref{Fig. S5}a). \label{Fig_Sx1}
}}
\end{figure*}

\section{Orbital projections of bands}
\label{sec:orbital}
Figure~\ref{Fig. S6} shows the orbital contributions to the bands for all the Ni 3d and S 3p levels separately. Other orbitals have negligible contributions in that energy range except for the lowest energy band (feature ix) that is dominated by contributions of P 3p$_z$ orbitals (Fig.~5a, main text).
All the bands down to $E-E_{\rm F}=-5.5$\,eV have a mixed character of Ni 3d and S 3p orbitals, while the displayed bands at lower energy  do not have any contributions from Ni orbitals.  
The topmost bands (feature i) have mostly S 3p character, but  significant contributions from Ni 3d$_{xz,yz}$. The features ii-vi are dominated by Ni 3d$_{z^2}$ and Ni 3d$_{xy,x^2-y^2}$ character, however, with varying contributions from the different S 3 p orbitals. The changing band across $T_{\rm N}$ features Ni 3d$_{z^2}$ and 3d$_{xy,x^2-y^2}$ as well as S 3p$_z$ contributions of similar strengths. 
A bunch of bands at $E-E_{\rm F}\le -5$\,eV  consists of Ni 3d$_{xz,yz}$, 3d$_{xy,x^2-y^2}$ and  S 3p$_{x,y}$ orbitals and, hence, is barely visible in the ARPES data via the simplified selection rules \cite{Moser2017}. These bands are faintly apparent at $h\nu=50-55$\,eV and $E-E_{\rm F}\approx -5$\,eV (Fig.~\ref{Fig. S5}c--d) due to a mixing with Ni 3d$_{z^2}$ (Fig.~\ref{Fig. S5}h--i). The bands between -6\,eV and -8\,eV have a nearly pure S 3p$_{x,y}$ character with minor contributions from P 3p$_{x,y}$ (not shown). In line, they appear only very faintly in the ARPES data at all $h\nu$.
For feature ix, most contributions originate from P 3p$_z$ orbitals with smaller contributions from the various S 3p orbitals. 

Most importantly, the energy range where bands change across $T_{\rm N}$, i.e. feature iv (and maybe vi), exhibits a strong mixing of all Ni 3d and S 3p levels implying a multi-orbital hybridization. 
This directly reflects the S-mediated superexchange between nearest neighbor Ni atoms  as discussed in the main text (Fig.~4, main text). 

%
\section{Additional charge density plots for states in the energy range changing with temperature}
\label{sec:CDPlot}

The identified energy range of the bands that shift with temperature (feature iv in Fig.~\ref{Fig. S6}a) contains four states at $\Gamma$ as marked by the arrow in  Fig.~3c, main text. The state that exhibits a charge density compatible with a nearest neighbor Ni 3t$_{\rm 2g}$-S 3p-Ni 3t$_{\rm 2g}$ superexchange path is plotted in Fig.~4, main text. The other three states are shown in Fig.~\ref{FigS7a}. None of them couples Ni atoms. Again the charge density is distinct in the ferromagnetically coupled zigzag chains of opposite spin direction. For the state in Fig.~\ref{FigS7a}b and e, the left zigzag chain ($\uparrow$), corresponding to the upper left Ni atom  in Fig.~\ref{FigS7a}e, shows a Ni 3t$_{\rm 2g}$ orbital. This orbital is coupled to one S 3p level via its top lobe. The bottom lobe  also couples to a S 3p-orbtal that itself is tilted towards a P-atom that is also connected to  two S 3p  lobes on the other side. In the other zigzag chain ($\downarrow$, upper right Ni atom in Fig.~\ref{FigS7a}e), the Ni atom  exhibits a nearly unperturbed 3t$_{\rm 2g}$ orbital. 

The state in Fig.~\ref{FigS7a}c and f shows uncoupled S 3p and Ni 3t$_{\rm 2g}$ orbitals in the left zigzag row (not visible in f). The two red in-plane lobes of the Ni 3t$_{\rm 2g}$ orbital are asymmetrically perturbed, but do not overlap with any neighbors also at larger extension of the contour planes (not shown). In contrast, the Ni 3t$_{\rm 2g}$ orbitals in the right zigzag row show one blue lobe each that couples relatively strongly to two neighboring S 3p orbitals. But again, this charge density is not connected to neighboring Ni atoms.  A minor charge density is also found between the two P atoms of the dumbbell. For the last state at this energy (Fig.~\ref{FigS7a}d and g), the Ni atoms in both rows show 3t$_{\rm 2g}$ orbitals interacting with neighboring S 3p orbitals. In the left zigzag row, the upper and the lower lobes exhibit overlap with an adjacent S 3p orbital. This is most clearly visible in Fig.~\ref{FigS7a}g for the Ni atoms with interacting red lobes at top and bottom, where both coupled S atoms are in front of the Ni atom. For the right zigzag row, one lobe of the Ni 3t$_{\rm 2g}$ orbital interacts with one and the other with two neighboring S 3p orbitals. But again, all these bonds are dead ends for connecting to neighboring Ni atoms.
Hence, none of the three states exhibits any exchange path, such that it is indeed likely that the state displayed in Fig.~4, main text, is the one that changes its energy across $T_{\rm N}$. 

 Interestingly, the same energy region for the Néel configuration (Fig.~3e, main text) shows a state that features a third nearest-neighbor exchange path  (Fig.~\ref{FigS7b}). The most left Ni 3t$_{\rm 2g}$ orbital has a touching point for its right, red lobe to two adjacent S 3p orbitals (red lobes as well). The same applies to the most right Ni 3t$_{\rm 2g}$ orbital concerning its left, blue lobe interacting with the blue lobes of two adjacent S 3p orbitals. These touching points of lobes evolve into a merging charge density  for more extended contour planes (not shown) corroborating its bonding nature. Intriguingly, the opposite lobes of the four involved S 3p orbitals are extended towards the intermediate P dumbbell such that they can interact pairwise at the position of one P atom as again corroborated by a charge density merger at more extended contour planes  (not shown). Hence, each of the central P atoms mediates a third nearest neighbor exchange of Ni atoms via Ni 3t$_{\rm 2g}$-S 3p-S3p-Ni 3t$_{\rm 2g}$ without involving its own orbitals. A similar configuration has recently been proposed to be responsible for the strong third nearest neighbor exchange interaction in NiPS$_3$ amounting to $J_3=-6.5$\,meV \cite{Lanon2018,Chittari2016}, however, involving two Ni 3e$_{\rm g}$ orbitals \cite{Scheie2023}. 

\begin{figure*}[!thbp]
\centering
\includegraphics[width=1\textwidth]{Fig_S10.pdf}
\vspace{-0.3 cm}
\caption{\textbf{Antiferromagnetic vs. paramagnetic band structure.}
\MM{(a) Spin arrangement of the atoms for the calculated antiferromagnetic (AFM) zigzag and the spin-disordered, paramagnetic (PM) state.
(b) Projected density of states (pDOS) of the AFM zigzag configuration for the three different elements (left)  and electronic band structure for both, the AFM zigzag and the PM state (right). Only states below the valence band maximum (0\,eV) are displayed. Dashed, horizontal line stresses the upwards shift of the bands marked by an arrow by roughly 150\,meV. (c) Same as b, but for states with $E>-1$\,eV. }\\ 
}
\label{Fig. S8}
\end{figure*}

 We checked also the charge distribution of other states in a broader energy range for both, the zigzag and the Néel configuration, but  without achieving any further reportable insight.

\section{Electronic band structure in $\overline{\rm K}\overline{\Gamma}\overline{\rm K}$ direction}
\label{sec:KGK}

Figure~\ref{Fig. S7} shows the comparison of the band structures in $\overline{\rm K}\overline{\Gamma}\overline{\rm K}$ and $\overline{\rm M}\overline{\Gamma}\overline{\rm M}$ direction, both for ARPES at $h\nu=60$\,eV and corresponding DFT+U data ($U_{\rm eff}=1.6$\,eV, $k_z=0.1$/\AA). 
The band structure in the two directions is rather similar in ARPES and DFT+U, with minor changes of $k_\parallel$ dispersion. Some similarities appear in the dispersion change between ARPES and DFT+U. For example, the low energy P 3p/S 3p band at $E-E_{\rm F}\approx -8.5$\,eV is more parabolic up to the BZ boundary in  $\overline{\Gamma {\rm K}}$ direction for ARPES and DFT+U. Moreover, comparing the dispersion around $\overline{{\rm K}}$ and $\overline{{\rm M}}$, the upwards dispersion towards $\overline{{\rm M}}$ at $E-E_{\rm F}\approx-4$\,eV is also visible in DFT+U, but in both cases not around $\overline{{\rm K}}$. Similarly, the upwards dispersion towards $\overline{{\rm K}}$ at  $E-E_{\rm F}\approx-5.5$\,eV, not apparent towards $\overline{{\rm M}}$ in ARPES, coincides with a more flat steepness along $\overline{\Gamma {\rm K}}$ in  DFT+U.


\begin{figure*}[tbh]
\centering
\includegraphics[width=1\textwidth]{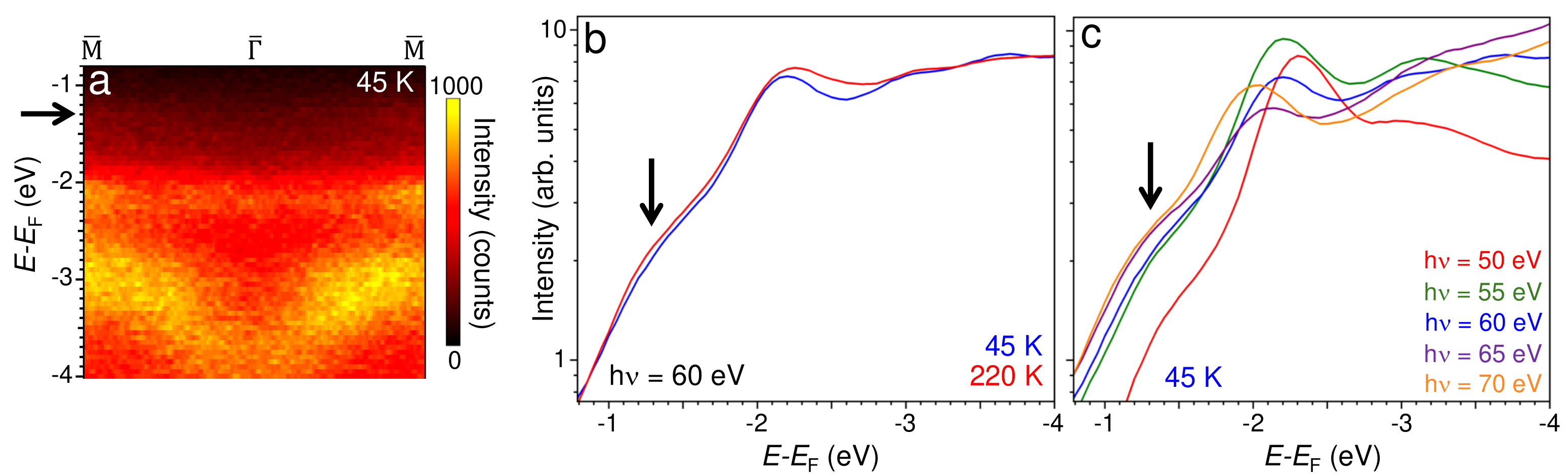}
\vspace{-0.0 cm}
\caption{\textbf{Spectral shoulder not explained by DFT+U.} (a) Raw ARPES data, $h\nu=60$\,eV, $T=45$\,K (Zoom into Fig. 2b, main text). (b) Smoothed energy distribution curves at $\overline{\Gamma}$ (box smoothing of two energy pixels (100 meV)), $T$ as marked, $h\nu=60$\,eV. (c) Same as b at various $h\nu$, $T=45$\,K. The arrow in a-c is consistently at $E-E_{\rm F}=-1.3$\,eV, i.\,e. at the ARPES feature for $h\nu=60$\,eV not covered by DFT+U calculations (Fig. 3c-d, main text).}
\label{Fig. S9}
\end{figure*}

\section{Comparison of DFT+U band structure along non-equivalent $\overline{\rm K}\overline{\Gamma}\overline{\rm K}$ directions}
\label{sec:directionalDFT}

\BB{Figure~\ref{Fig_Sx1} compares the DFT+U band structures along the three $\overline{\rm K}\overline{\Gamma}\overline{\rm K}$ directions in the hexagonal BZ. These directions are non-equivalent
in the magnetic zigzag configuration via the resulting rectangular BZ (Fig.~\ref{Fig. S1}a). Indeed the band structures exhibit minor differences as highlighted by arrows in Fig.~\ref{Fig_Sx1}b-d, but they are well below 100\,meV or burried into bunches of bands. Hence, they are rather challenging to be observed in our experiment with energy resolution of 50\,meV and typical peak widths above 200\,meV (Fig.~2c,f, main text, Suppl. Fig.~\ref{Fig. S9}b-c).  Moreover, since the 3D BZ is not mirror symmetric perpendicular to $k_z$ except at $k_z=0/$\AA\, (Fig.~\ref{Fig. S5}a), the projected directions as probed by ARPES are already non-equivalent without magnetic symmetry breaking. Most strikingly, the positive and negative $\overline{\Gamma}\overline{\rm K}$ directions become non-equivalent by the 3D BZ shape leading to distinct not point-symmetric dispersions (Fig.~\ref{Fig_Sx1}e-g) as partially also apparent in the experimental data (Fig.~\ref{Fig. S5}b-g). Both complicates the detection of magnetically induced symmetry breakings additionally. }


\section{Comparing zigzag and disordered magnetic configurations}

Figure~\ref{Fig. S8} compares the band structure of NiPS$_3$ for the zigzag AFM ground state and a randomly disordered  configuration of atomic spins in a $4\times 2\times 3$ supercell with zero net magnetic momentum in \MM{the supercell (Fig.~\ref{Fig. S8}a)}. The latter is a reasonable approximation of the paramagnetic case \cite{Trimarchi2018, Strasdas2023}. In both cases, we unfold to the atomic, hexagonal surface BZ (Fig.~\ref{Fig. S5}a). Here, we only display the intensity of the bands after unfolding \MM{(Fig.~\ref{Fig. S8}b, c), since a projection to individual orbitals is rather challenging after unfolding the extremely small BZ of the disordered configuration. For assignment, we add the element specific partial density of states (pDOS) of the zigzag AFM configuration (left to the band structures).}

\begin{figure}[htb]
\centering
\includegraphics[width=0.5\textwidth]{Fig_S12.pdf}
\vspace{-0.0 cm}
\caption{\textbf{ARPES data above and below $T_{\rm N}$ compared to DFT+U data.} \BB{ (a) ARPES curvature along $\overline{\rm M}\overline{\Gamma}\overline{\rm M}$ direction, $h\nu=60$\,eV, $T=220$\,K. (b) ARPES curvature along $\overline{\rm M}\overline{\Gamma}\overline{\rm M}$ direction, $h\nu=60$\,eV, $T=45$\,K. (c) Calculated band structure by DFT+U, $U_{\rm eff}=1.6$\,eV, \textit{k}$_\textit{z} = 0.1$/\AA, only bands with significant s, p$_z$ and d$_{z^2}$ character are displayed \cite{Moser2017}, same as Fig.~\MM{\ref{Fig. S2}d}.}}
\label{Fig. S12}
\end{figure}

Comparing the band structures (Fig.~\ref{Fig. S8}b--c), one observes large similarities including a barely changed \MM{fundamental} band gap at $E=0-1$\,eV. In the energy range with significant Ni 3d contribution of both spin directions, the bands are more blurred in the paramagnetic phase. This is likely due to the different energies of various Ni atoms  due to  their different local spin environment.
Notice that the contribution of each Ni atom is projected to the same $k_\parallel$ via the unfolding, since the Ni atoms with different spin  are forced into the corresponding periodicity of the smaller unit cell. Consistently, bands with pure S or P character are barely blurred since they are largely periodic in the smaller unit cell. 

Importantly, the most striking band change  appears at the crossing point (at $\overline{\Gamma}$, $E\approx -2$\,eV, white arrows in Fig.~\ref{Fig. S8}b) that we identified as the region of band shift in the ARPES experiment.
The lower part of the crossing moves upwards \MM{by about 150\,meV} (dashed line)  and broadens in the paramagnetic phase. This corroborates that this energy region is indeed very sensitive to the magnetic configuration, albeit the changes could not be reproduced quantitatively via this simplified version of a paramagnetic state.


Nevertheless, the favorable comparison demonstrates that modeling the paramagnetic phase by spin disorder reveals important insights, in that case corroborating the assignment of changing bands in the experiment. 
This is good news regarding the extensive  computational cost of finite temperature calculations, in particular for large unit cells and significant correlations as in the case of NiPS$_3$ \cite{Trimarchi2018}. 

\section{ARPES Feature not covered by DFT+U calculation}

Figure~\ref{Fig. S9} shows the additional feature observed at $E-E_{\rm F}\approx - 1.3$\,eV in more detail. The raw $I(E,k_\parallel)$ plot reveals a sizable contrast change at the corresponding energy (Fig.~\ref{Fig. S9}a, arrow) implying that it is a real structure. Energy distribution curves at $\Gamma$ show the feature as a shoulder below the peak of the valence band maximum at $E-E_{\rm F}\approx-2.2$\,eV (Fig.~\ref{Fig. S9}b). The energy position and height of the shoulder are nearly identical above and below $T_{\rm N}$. However, the shoulder shifts upwards in energy with increasing $h\nu$ very similarly to the valence band maximum (Fig.~\ref{Fig. S9}c). This parallel upwards shift can be seen more clearly in the curvature plots of Fig.~\ref{Fig. S5}c-g, where both barely dispersive features move upwards by roughly $ 200$\,meV in parallel between $h\nu=50$\,eV and $h\nu=70$\,eV. Hence, the shoulder is robust and appears to be tied to the valence band maximum. As mentioned in the main text, its origin will be discussed in a separate publication.

\section{Comparison of the full energy range of the ARPES data above and below T$_N$.}

\MM{Fig.~\ref{Fig. S12} shows the full energy range of the ARPES data recorded above and below $T_{\rm N}$ in comparison with the optimized DFT+U data of the band structure. For the latter, we again display only the bands that are recorded according to the simplified selection rules \cite{Moser2017}. One recognizes that changes only appear in the energy range of $E-E_{\rm F}=-3$ to $-4$\,eV. In particular, the assignement of the calculated band structure via $U_{\rm eff}$ (Fig.~\ref{Fig. S2}) and $k_z$ (Fig.~\ref{Fig. S4}) is not influenced by these minor changes. Hence, the calculated band structure fits nicely to the ARPES data above and below $T_{\rm N}$ except of the mysterious structure at $E-E_{\rm F}=-1.3$\,eV. }

\section{ARPES curvature}

\ML{In the main text as well as in the supplementary information, we partially use ARPES curvature plots. This enables for better visibility of the bands \cite{Zhang2011}. The curvature $C(E)$ is calculated with respect to energy $E$ after smoothing the raw data, as described in the caption of Fig.~2c, main text, via:   
\begin{equation}
 C(E) = \frac{-I''(E)}{(C_0 + I'(E)^2)^{3/2}}\,\hspace{2mm}{\rm with}\,\hspace{2mm} C_0= a_0 |I'(E)|_\text{max}^2.
 \label{eq:Scurv1}
\end{equation}}
 
\ML{Here, $I'$ and $I''$ are the first and second partial derivative with respect to energy of the smoothed photoelectron intensity $I = I(E,k_\parallel)$, $|I'(E)|_\text{max}$ is the maximum value of the first partial derivative, while $a_0$ is a free parameter used to adjust the relative intensity of different bands for better visibility. We have consistently used $a_0=0.05$.}

\ML{To highlight the advantage of using curvature instead of the second derivative, we shortly describe its origin.
The curvature $C(x)$ of a function $y:=f(x)$ measures the change of angle $\varphi$ of a graph's tangent with respect to the path length $s$ along the graph. This reads
\begin{equation}\label{eq:Scurv2}
    C(x) = - \frac{d\varphi}{ds} = - \frac{d\varphi}{dx} \frac{dx}{ds} = - \frac{f''(x)}{\left( 1+\left( f'(x) \right)^2 \right)^\frac{3}{2}}
\end{equation}
with $\tan \varphi(x) = \frac{dy}{dx} = f'(x)$ \cite{Tsai1994}. The negative sign is chosen to identify relative maxima of $f(x)$ with the maxima of $C(x)$.
$C(x)$ features the local inverse radius $1/R$ of the $y(x)$ plot, but only if $x$ and $y$ are equally scaled and, hence, have the same unit. For example, a full circle then implies directly $d\varphi/ds = 2 \pi / (2 \pi R) = 1/R$. Typically, the smallest tangential circles are found at the local maxima of a curve (Fig.\ \ref{Fig. S13}a). Consistently, a local maximum in $f(x)$ implies that  $f'(x)$ vanishes while $-f''(x)$ has a large, positive value, such that the corresponding $C(x)$ mostly exhibits a local maximum, too. Importantly, the curvature showcases the inverted parabola area around a peak more directly than the second derivative. If one considers a purely inverted parabola $y=-x^2+4$, it exhibits a peak in curvature at its maximum $\left( C(x) = \frac{2}{\left( 1 + 4x^2 \right)^{3/2}} \right)$, while the $2^\mathrm{nd}$ derivative remains constant (Fig.\ \ref{Fig. S13}b).} 

\ML{These favorable properties of the curvature persist, if $y$ and $x$ have different units as in the $I(E)$ curves of the ARPES data. However, a local radius of a curve is then not well defined due to the unit mismatch. Hence, the 1 in the sum of the denominator of eq.\ (\ref{eq:Scurv2}) gets arbitrary, besides having the wrong unit \cite{Zhang2011}.
Consequently, $C_0$ is introduced in eq.\ (\ref{eq:Scurv1}) that can be used as a tuning parameter balancing between the first and the second derivative as the two benchmarks of a local peak position.}  

\ML{Reducing $C_0$ strengthens real maxima by the divergence of $1/f'(x)^3$ (Fig.\ \ref{Fig. S13}b). In contrast, increasing $C_0$ strengthens the $-f''(x)$ term and, hence, highlights shoulders in the ARPES data that originate from peaks on top of a background intensity. The optimal value of $C_0$ has to be chosen visually by careful comparison to the raw $I(E)$ data such that reproducible shoulders as in Fig.\ \ref{Fig. S9}b-c appear as features in $C(E)$ while apparent noise does not.}




\begin{figure}[htb]
\centering
\includegraphics[width=0.5\textwidth]{FigS13ab.pdf}
\vspace{-0.8 cm}
\caption{\ML{\textbf{Curvature versus second derivative} (a) Exemplary function $f(x) = -x^2+4$ with tangential circle at $f(x=0)$. The radius $R=0.5$ is marked in red. (b) The same $f(x)$ as in (a) (blue) is plotted together with the $2^\mathrm{nd}$ derivative $d^2f(x)/dx^2$ (orange) and curvature $C(x)$ according to eq.\ (\ref{eq:Scurv1}) with two different parameters $C_0$ (yellow, violet).}}
\label{Fig. S13}
\end{figure}

\section{Comparison between DFT+U, hybrid functionals and GW calculations}

\Magda{In this section, we clarify our rationale for using the DFT+U framework. We support our choice with additional calculations using hybrid functionals and post-DFT methods such as GW.}

\Magda{The effective Hubbard parameter $U_{\rm eff}$ is not a universal constant, but rather depends strongly on the specific modeling framework. Its numerical value is sensitive to the choice of exchange–correlation functional, the inclusion of van der Waals corrections, and the type of projector or localized basis used for defining the correlated orbitals. As a consequence, reported values of $U_{\rm eff}$ in the literature span a wide range, even for nominally similar materials.  For example, in the literature on NiO, values between 4\,eV and 8\,eV are commonly employed \cite{Rohrbach2005,Cococcioni2005,Panda2016,Kunes2007}, whereas for layered chalcogenides such as TlNi$_2$Se$_2$,  smaller values (2–3\,eV) can be found \cite{PhysRevB.108.045140}.   
 Importantly, our selection of $U_{\rm eff} = 1.6$\,eV is not arbitrary, but is guided by its ability to largely reproduce the details of the valence-band structure as observed in ARPES, which is central to the interpretation of the electronic structure presented in this manuscript.}  

\begin{figure}[!h]
\centering
\includegraphics[width=0.5\textwidth]{S14.png}
\vspace{-0.3 cm}
\caption{\Magda{Comparison between the orbital-projected, spin-polarized density of states (DOS) calculated using DFT+U ($U_{\rm eff}$=1.6 eV) and various exchange–correlation functionals as HSE06, HSE03, and PBE0, red: Ni 3e$_{\rm g}$-, black:  Ni 3t$_{\rm 2g}$-, grey: ligand orbitals.  Both, HSE03 and HSE06 employ an exact-exchange fraction of 0.25. Notice that the gap in HSE03, HSE06 and PBE0 is significantly larger than experimentally observed: $E_{\rm g}=1.8$\,eV \cite{Kim2018}.}}
\label{S14}
\end{figure}

\begin{figure*}[]
\centering
\includegraphics[width=0.9\textwidth]{S15.png}
\vspace{-0.3 cm}
\caption{\Magda{Comparison between the  orbital-projected, spin-polarized DOS calculated using  DFT+U ($U_{\rm eff}=1.6$\,eV) and hybrid functionals with different fractions of the exact exchange.}}
\label{S15}
\end{figure*}

Additionally, it is well established that experimental estimates of effective Coulomb interactions are themselves method-dependent. Different spectroscopic techniques (e.g., XPS, EELS, ARPES) probe distinct excitations and screening environments and therefore often yield different effective interaction strengths. This variation does not indicate a shortcoming of DFT+U, but rather stems from the fundamental difference between the quantities accessible in experiment---typically related to excited states---and those described by DFT, which is a ground-state theory. In principle, there exists a single value of $U_{\mathrm{eff}}$ that would correctly complement the exchange--correlation functional to reproduce both ground-state and excitation properties. However, this value is generally not known \textit{a priori}, and different attempts to extract it from experiment or theory can yield different results depending on how closely the measured excitation reflects the underlying ground-state properties. Importantly, no existing electronic-structure formalism---whether DFT+U, hybrid functionals, or GW---can completely eliminate this ambiguity, as each relies on different approximations and levels of treatment of electronic screening. 

 \Magda{To compare our calculations with alternative approaches to the exchange-correlation energies, we firstly present the comparison between the DFT+U with hybrid functionals  such as HSE06 \cite{10.1063/1.2404663}, HSE03 \cite{10.1063/1.1564060}, or PBE0 \cite{10.1063/1.472933} (Fig.~\ref{S14}). All hybrid functional calculations employ the optimized  geometry  within the PBE+U framework (with $U = 1.6$\,eV) including D3 van der Waals corrections. 
 It is well established that the critical parameter in such functionals is the fraction of the Hartree–Fock exchange. It has a direct and significant impact on the relative energies of the correlated 3d states (Fig.~\ref{S15}).  The default fraction of 25\% of Hartree-Fock exchange systematically overlocalizes the Ni 3d orbitals. This leads to a disagreement with ARPES measurements, in particular, the pronounced Ni 3e$_{\rm g}$ contribution at $E\in[-5\,{\rm eV},-6\,{\rm eV}]$ is not found by ARPES. Reducing the fraction of the Hartree-Fock exchange restores the agreement with the DFT+U at $U_{\rm eff}=1.6$\,eV (Fig.~\ref{S15}) and, hence, with the ARPES data, but this amounts to the same type of empirical adjustment that one performs with $U_{\rm eff}$ in DFT+U.  In particular, with 10\% of Hartree-Fock exchange, one reproduces the  DFT+U results at the chosen $U_{\rm eff}=1.6$\,eV (Fig.~\ref{S15}), while higher percentages reveal the same trends as larger $U_{\rm eff}$ in DFT+U (not shown, but compare to Fig.~\ref{Fig. S2}b-g). This illustrates the close analogy between tuning the fraction of Hartree–Fock exchange and adjusting the Hubbard $U_{\rm eff}$ parameter. Importantly, no additional band is appearing above the valence band maximum, while only the Ni 3d levels are shifted downwards in the valence band and upwards in the conduction band with increasing percentage of the exact exchange contribution (Fig.~\ref{S15}).}

\begin{figure*}[]
\centering
\includegraphics[width=0.75\textwidth]{S16.png}
\vspace{-0.3 cm}
\caption{\Magda{Comparison between the  orbital-projected, spin-polarized DOS using DFT+U ($U_{\rm eff}=1.6$\,eV) as well as calculations with quasiparticle corrections to the Kohn-Sham density of states using  G0W0  and partially self-consistent EVGW0 calculations on top of PBE.}}
\label{S16}
\end{figure*}

This highlights that hybrid functionals are not entirely parameter-free when applied to materials with correlated electrons. In practice, they often require empirical adjustments, e.\,g., to reproduce experimental band gaps or other key properties, conceptually analogous to the tuning of the $U$ parameter in DFT+U. However, while $U$ acts directly on localized d-states, hybrid functionals affect all electronic states through the inclusion of exact exchange. For this reason, we consciously decided not to employ hybrid functionals in this study. Although their use would, in principle, be technically feasible, their computational cost renders them impractical for the large-scale simulations required here (e.g., performing Brillouin zone unfolding).

\Magda{We also carried out G0W0 \cite{PhysRevB.74.035101} and partially self-consistent EVGW0 \cite{PhysRevB.34.5390} calculations using the optimized structure calculated within  PBE+U framework ($U = 1.6$\,eV) including D3 van der Waals corrections (Fig.~\ref{S16}). For the GW calculations, we employed PAW pseudopotentials adapted for GW computations. In the GW implementation, we used a fully frequency-dependent approach without the plasmon-pole approximation. The number of bands was set to approximately ten times the number of occupied bands for both the G0W0 and EVGW0 approaches. For the EVGW0 method, four self-consistency (scfGW) iterations were performed to ensure convergence. These approaches provide a rigorous quasiparticle correction to the Kohn-Sham band structure, but their effect strongly depends on the starting point. Namely, in our case, they do not lead to significant changes in the ordering of the t$_{\rm 2g}$  and e$_{\rm g}$ manifolds, but just to an increase of the \MM{fundamental} band gap and a relatively small upwards shift of all bands (Fig.~\ref{S16}). Moreover, they do not exhibit an additional band feature above the valence band maximum such as our ARPES feature at $E-E_{\rm F}=-1.3$\,eV. Instead, a slight narrowing of the prominent Ni 3d peaks is found (Fig.~\ref{S16}).
These observations are consistent with the well-known sensitivity of GW to the underlying mean-field reference. We would like to stress that GW methods are extremely powerful in addressing excitation-related phenomena, such as band-gap renormalization or excitonic effects, and they have been widely applied to such problems. Their main advantage, however, lies in the treatment of quasiparticle excitations rather than in correcting the relative position of strongly localized 3d orbitals in transition-metal chalkegonides. For the latter, DFT+U remains a simpler and more robust choice.}

\Magda{In summary, our extended benchmark calculations confirm that neither hybrid functionals nor GW approaches capture the additional ARPES feature at $E-E_{\rm F}=-1.3$\,eV. The discrepancy is therefore not attributable to the choice of DFT+U. Importantly, DFT+U offers unique advantages that make it the most appropriate method for this study. Namely, it allows for a computationally tractable description of the paramagnetic state and the unfolding of electronic structures into a different supercell as required for comparison with the ARPES data. This would not be feasible with hybrid functionals or GW due to their computational cost. Furthermore, a comparison of NiPS$_3$ with related compounds such as MnPS$_3$ (Fig.~5, main text), where DFT+U was successful in capturing all bands found by ARPES,  indicates that one generally should not expect additional bands in the electronic structure beyond those reproduced by DFT+U. Any further splitting of ligand-derived states or localized $d$-manifolds would in fact be surprising given the crystal symmetry of this class of materials. This strongly suggests that the additional feature observed in ARPES above the valence-band maximum cannot be rationalized within a mean-field framework, but rather reflects physics that goes beyond standard DFT-based methods.}


\FloatBarrier
\bibliography{references}